\documentclass{article}

\usepackage{PRIMEarxiv}

\usepackage[utf8]{inputenc} 
\usepackage[T1]{fontenc}    
\usepackage{hyperref}       
\usepackage{url}            
\usepackage{booktabs}       
\usepackage{amsfonts}       
\usepackage{nicefrac}       
\usepackage{microtype}      
\usepackage{lipsum}
\usepackage{fancyhdr}       
\usepackage{graphicx}       
\graphicspath{{media/}}     

\usepackage{cite}
\usepackage{bm}
\usepackage{amssymb,amsmath}
\usepackage[justification=justified]{caption}
\usepackage[justification=justified]{subcaption}
\newcommand{\mi}{{\rm i}}
\newcommand{\md}{{\rm d}}
\newcommand{\me}{{\rm e}}
\newcommand{\rr}{{\bm r}}
\newcommand{\rrM}{{\bm r}_{\rm M}}
\newcommand{\rrA}{{\bm r}_{\rm A}}
\newcommand{\rrB}{{\bm r}_{\rm B}}
\newcommand{\ee}{{\bm e}}
\newcommand{\kk}{{\bm k}}
\newcommand{\GG}{{\bf G}}
\newcommand{\GGb}{{\bf G}^{(0)}}
\newcommand{\GGp}{{\bf G}^{(1)}_{\rm pl}}
\newcommand{\GGs}{{\bf G}^{(1)}_{\rm sp}}
\newcommand{\IM}{{\rm{Im}}\,}

\newcommand{\stat}[1]{\left|#1\right\rangle}

\title{Purcell-induced suppression of superradiance for molecular overlayers on noble atom surfaces
\thanks{\textit{\underline{Citation}}: 
\textbf{J. Fiedler, K. Berland, S.Y. Buhmann. Purcell-induced suppression of superradiance for molecular overlayers on noble atom surfaces.  DOI:10.1063/5.0106503.}} 
}

\author{
  Johannes Fiedler\thanks{Physikalisches Institut,
 Albert-Ludwigs-Universit{\"a}t Freiburg,
 Hermann-Herder-Str. 3, 79104 Freiburg, Germany} \\
 Department of Physics and Technology\\
 University of Bergen\\
 All\'egaten 55, 5007 Bergen, Norway\\
  \texttt{johannes.fiedler@uib.no}\\ 
   \AND
  Kristian Berland \\
Department of Mechanical Engineering\\ and Technology Management\\ Norwegian University of Life Sciences\\ Campus {\AA}s Universitetstunet 3, 1430 {\AA}s, Norway \\
 \And
  Stefan Y. Buhmann\\
  Institut f\"ur Physik, Universit\"at Kassel\\
  Heinrich-Plett-Str. 40, 34132 Kassel, Germany\\
}

\begin{document}
\maketitle

\begin{abstract}
We study the impact of an environment on the electromagnetic responses of a molecule in the presence of a dielectric medium. By applying the dipole--dipole coupling between the molecule's and the environment's degrees of freedom, we can reduce the complex system into its components and predict excitation lifetimes of single and few molecules attached to a dielectric surface by knowing the entire quantum-mechanical properties of the molecules, such as transition energies and dipole moments. The derived theory allows for the description of superradiance between two molecules depending on the geometric arrangement between both concerning their separation and orientation with respect to each other. We analyse the possibility of superradiance between two molecules bound to a dielectric sphere and determine a change of the relevant length scale where the usually considered wavelength in free space is replaced with the binding distance, drastically reducing the length scales at which collective effects can take place.
\end{abstract}


\section{Introduction}
Organic photovoltaics and optoelectric devices, in which optical properties arise from the response of molecular complexes, aggregates or nanostructures, have attracted much attention.~\cite{nano10112192,PhysRevLett.123.163202,Christou2021,PhysRevApplied.11.024009}. Recent investigations on the photoexcitation-driven processes in such systems~\cite{doi:10.1021/acs.jpclett.7b00319,Stienkemeier_2006} aim to increase of the efficiency of organic solar cells~\cite{moraitisNanoparticlesLuminescentSolar2018,nayakPhotovoltaicSolarCell2019}, for instance by exploiting concepts such as singlet fission. For the optical properties of molecular complexes and nanostructures, two effects play an important role in understanding how the response of an individual molecule differs from that of the full system. One is the Purcell effect~\cite{Fuchs2018,Purcell1946}, i.e., the environmental influence on the decay rate of an excited state, which is a single-molecule effect in an effective environment. In contrast, the corresponding collective effects of an ensemble of molecules yields superradiance~\cite{Dicke1954,Gross1982,PhysRevLett.119.097402}.

Depositing molecules on noble gas surfaces is an attractive strategy to study the molecule's properties, as the noble gas surface only weakly perturbs many properties of molecules.~\cite{Stienkemeier_2006}  In this paper, we investigate whether adsorption on noble gas surfaces is a viable strategy for studying the superradiance of molecules.
Our investigations indicate that even for weakly interacting noble gas surfaces, such as Neon, superradiance would only occur for separations so short that typical molecules with a strong optical signal would overlap, entirely changing the nature of the system. 
Thus, the presence of a weakly interacting surface dramatically suppresses superradiance compared to a free-standing monolayer.

\begin{figure}
    \centering
    \includegraphics[width=0.4\columnwidth]{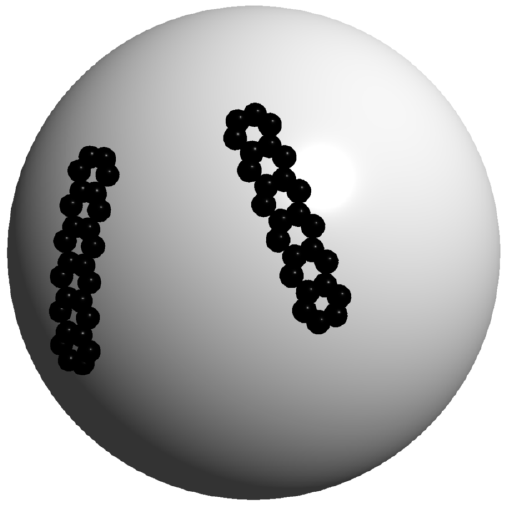}
    \caption{Schematic figure of the considered system: two molecules (here: pentacene molecules) are attached via van der Waals forces to a spherical nanoparticle with radius $R$.}
    \label{fig:scheme}
\end{figure}

Figure~\ref{fig:scheme} depicted the considered scenario. We assume that two molecules are weakly bound to a dielectric sphere with radius $R$ via Van der Waals forces. The Van der Waals assumption allows the separation of the system into three interacting subsystems due to the spatial separation of the electronic wave functions of each subsystem~\cite{D0CP02863K}. We calculate the sphere's impact on the single molecule's properties by perturbing its wave functions due to the presence of the dielectric object in dipole approximation, leading to point particles. Furthermore, we consider the impact of a second molecule on the optical local mode density, which can be generalised to $N$ interacting molecules and allows for the description of superradiance in such systems. Current theoretical studies address superradiance by simulating the entire system~\cite{PhysRevLett.119.097402} restricting the observed effects to a fixed molecule density that is usually assumed to be dense. We also consider a dilute packing of the molecules on the substrate, dramatically influencing the superradiance properties. To illustrate the model, we will explicitly consider the superradiance of the $S_1\rightarrow S_0$ transition of Acene near Argon or Neon clusters; see the experiments reported in Ref.~\cite{Matthias21}

Superradiance is a collective effect of a group of emitters interacting via an electromagnetic field. If the wavelength of the field is much larger than the emitter's separations, the ensemble will interact with the field collectively and coherently~\cite{Gross1982,Fuchs2018,PhysRevA.71.011804,PhysRevResearch.3.L032016}. Thus, the collective state reads as
\begin{equation}
\stat{I} = \frac{1}{\sqrt{N}} \biggl(\stat{1,0,0,\dots} + \stat{0,1,0,\dots}+\dots\biggr)\,,\label{eq:init}
\end{equation}
meaning that all particles share a single excitation. Furthermore, we consider both particles to be of the same species and located at the same distance to the interface. Thus, we can neglect the impact of surface-induced state detunings  because they occur equally for both particles~\cite{Ribeiro2015}. To observe this effect experimentally, the superradiant eigenstate of the dimer (the combined system of both molecules) has to be of lower energy than the subradiant one. Otherwise, the excitation will most likely decay non radiatively.
In terms of quantum optics, this effect goes hand in hand with an increase in the transition rate $\Gamma$, which is proportional to the contraction of the Greens tensor with the molecular transition dipole moments
\begin{equation}
    \Gamma \propto {\bm{d}}^\star \cdot {\rm{Im}}\,\GG\cdot {\bm{d}}\,.
\end{equation}
The exact formula is given in Eq.~(\ref{eq:Gamma}). The geometric arrangement is encoded in the scattering Green function. To this end, superradiance originates in the enhancement of the local mode density due to the presence of the other molecules~\cite{Scheel2008}, which is proportional to the imaginary part of the scattering Green tensor. By considering two molecules at position $\rrA$ and $\rrB$, we can write the field enhancement proportional to the sum over the Green functions
\begin{eqnarray}
    &&\IM g(\rrA,\rrA) + \IM g(\rrB,\rrB) \nonumber\\&&+ \IM g(\rrA,\rrB)+\IM g(\rrB,\rrA) \,,\label{eq:balance}
\end{eqnarray}
induced by each molecule [equal positions $\IM g(\rr,\rr)$: local-mode density] and a cavity-like resonance [different positions $\IM g(\rr,\rr')$: collective atom-field coupling constants], where we assumed the in-phase combination of the sheared excitation~(\ref{eq:init}). The scalar Green function $g(\rr,\rr')$ represents the relevant direction of the dyadic Green function due to the orientation of both molecules, which is its perpendicular component in the presence of a surface or its trace in free space. The first two terms in Eq.~(\ref{eq:balance}) denote the mode density of each molecule and the cross-terms leading to the superradiance enhancement. Thus, we can obtain a measure by comparing the total local mode density to the one generated by the single molecules as a superradiance fidelity
\begin{eqnarray}
\sigma= 1+\frac{\IM  g(\rrA,\rrB)+\IM g(\rrB,\rrA)}{\IM g(\rrA,\rrA)+\IM g(\rrB,\rrB)}\,,\label{eq:sigma}
\end{eqnarray}
which indicates superradiance for values close to two, $\sigma\lesssim 2$ and the absence of superradiance for values close to unity. 
\begin{figure}[t]
    \centering
    \includegraphics[width=0.8\columnwidth]{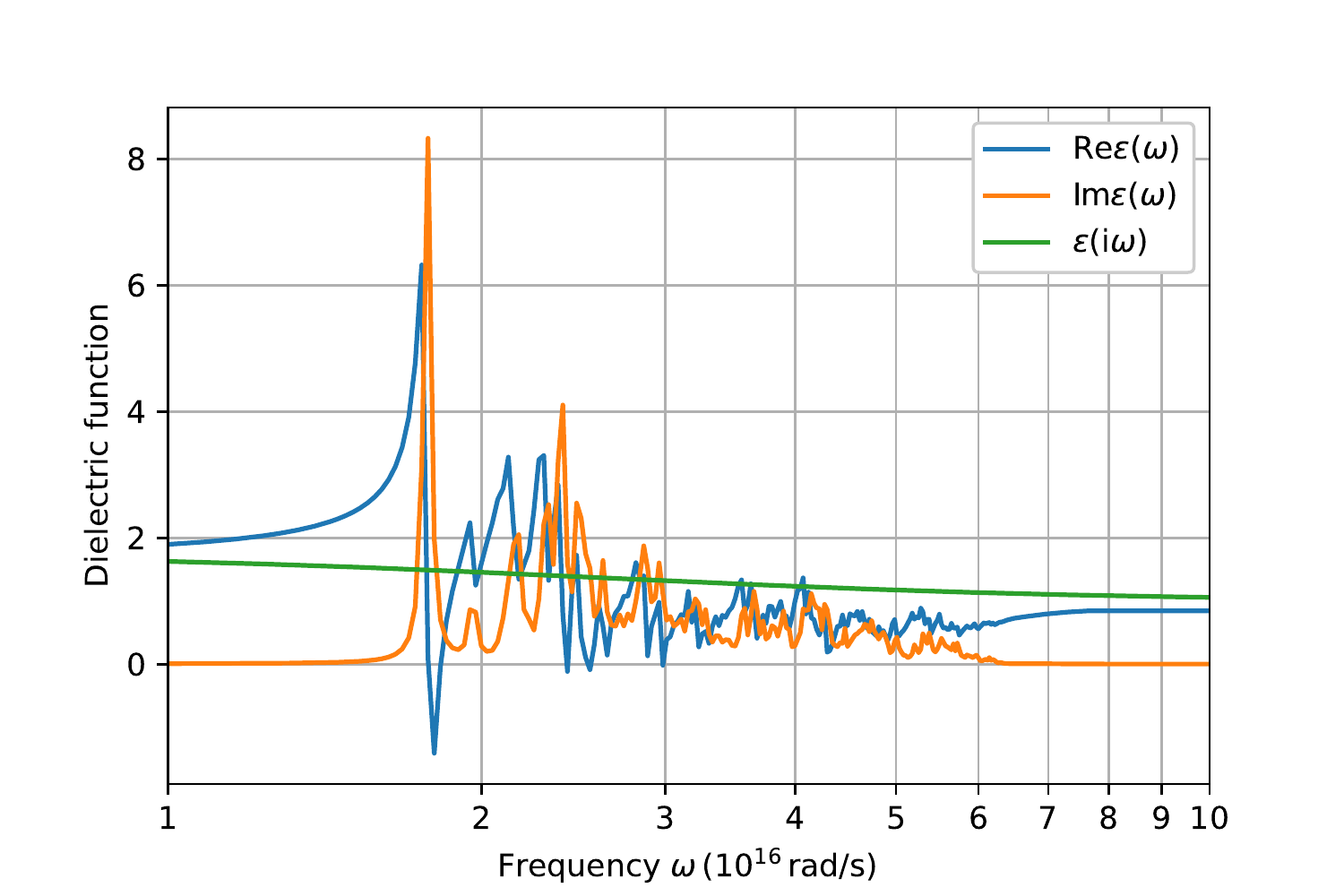}
    \caption{Dielectric function of solid Argon: real part (blue curve), imaginary part (orange), and on imaginary frequency axis (green).}
    \label{fig:Argon}
\end{figure}
\section{Dielectric function of Argon and Neon}\label{sec:DFT}
The dielectric function of the Argon crystal was computed using the many-body perturbation theory based on the Bethe--Salpether equation (BSE) in the Tamm--Dancoff approximation.\cite{Rohlfing1998} One-molecule excitation energies are obtained at the G0W0 level based orbitals obtained via density functional theory (DFT) with the PBE0-1/3 functional.~\cite{pbe0_one_third} Experimental lattice constants are used~\cite{argon_crystal}. All the calculations are performed with the \textsc{VASP} software package.~\cite{vasp1,vasp3,vasp4} In the calculations, we use a $6\times 6\times 6$ sampling of the Brillouin zone with 128 electronic bands for the G0W0 calculations. We use 15 virtual occupied and 15 unoccupied band orbitals in the BSE calculations. The GW-BSE method is well established as an accurate method for obtaining optical spectrum, which is confirmed by the fact that the static dielectric constant of $\epsilon_1(0) = 1.71$ of Argon agrees well with the experimental value of 1.67.~\cite{Sinnock_1980} Moreover, the first peak position at 11.67~eV is in good agreement with the experimental peak position at 12.06 eV.~\cite{Saile1976}

\begin{figure}[t]
    \centering
    \includegraphics[width=0.8\columnwidth]{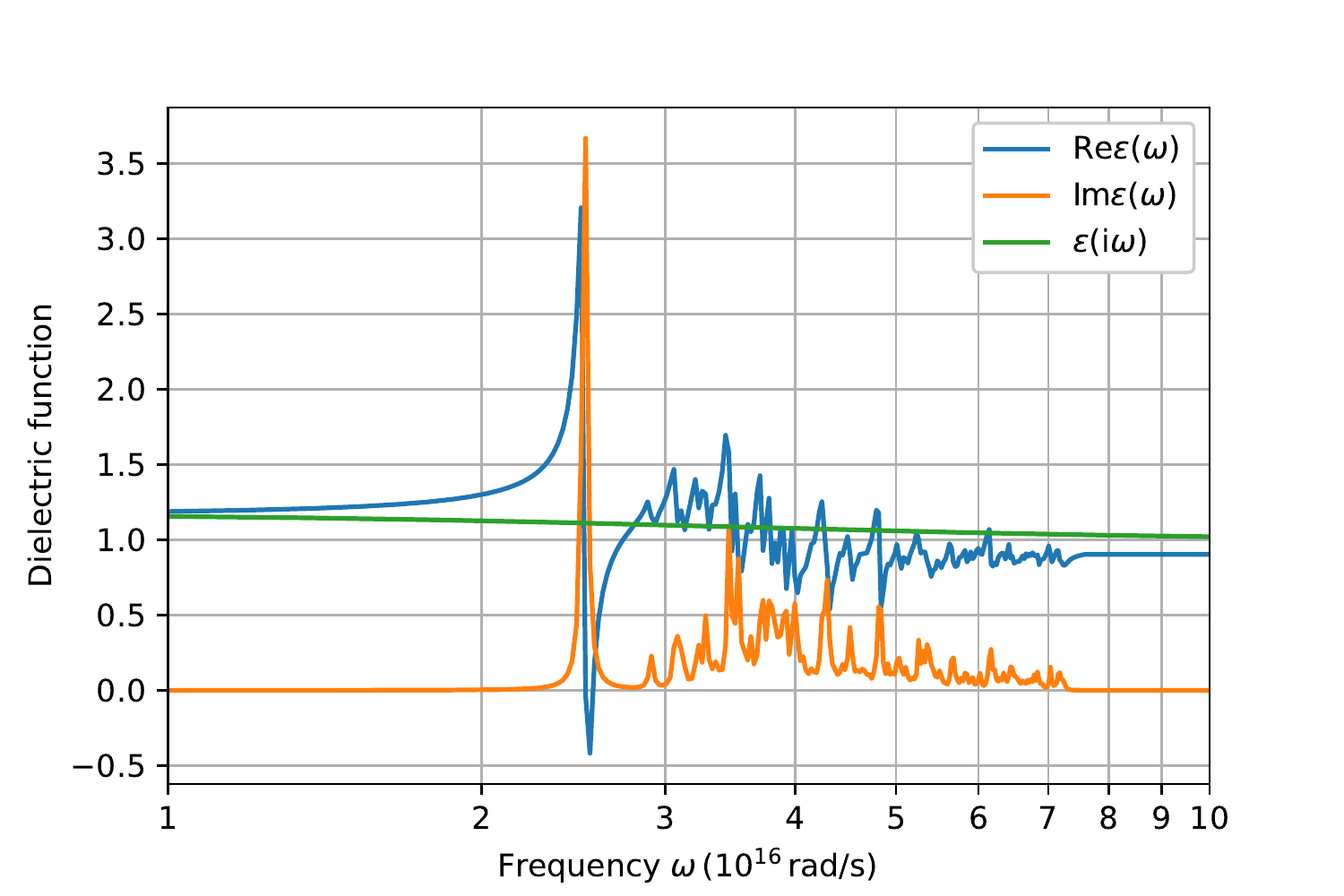}
    \caption{Dielectric function of solid Neon: real part (blue curve), imaginary part (orange), and on imaginary frequency axis (green).}
    \label{fig:Neon}
\end{figure}

\section{Theoretical model}
The dependence of the excitation lifetime on the separation between the pentacene molecule and the argon cluster and on the size of the cluster itself is obtainable via the internal dynamics of the molecule's energy levels by coupling to the ground-state electromagnetic field. This approach is based on macroscopic quantum electrodynamics.~\cite{Dung2003,Buhmann2004} 
In this theory, a molecule with discrete energy levels is coupled to the local mode density, which is enhanced by the presence of the cluster. We outline the fundamental steps in deriving the theory and applying spherical geometries. We make quantitative predictions of the excitation lifetime using the formalism with material and response properties obtained with density functional theory.

\subsection{Internal dynamics of molecules near interfaces}
The excitation lifetime describes the dynamics of a decay process within the molecule. Thus, one refers to such processes as internal dynamics within the quantum-optical framework. 
To describe the internal quantum mechanical dynamics of a molecule in the presence of dielectric bodies, we obtain the Hamiltonian by separation into the subsystems: the molecular system $\hat{H}_{\rm M}$; the electromagnetic fields $\hat{H}_{\rm F}$; and the molecule-field coupling $\hat{H}_{\rm MF}$,
\begin{equation}
    \hat{H} = \hat{H}_{\rm M} + \hat{H}_{\rm F} +\hat{H}_{\rm MF}\,. \label{eq:Ham}
\end{equation}
Molecular states $\left|n\right\rangle$ are described by an infinite set of discrete wave functions leading to the diagonalised molecular Hamiltonian
\begin{equation}
    \hat{H}_{\rm M} = \sum_n E_n \left|n\right\rangle\left\langle n\right| =\sum_n E_n \hat{A}_{nn}\, ,\label{eq:HamAtom}
\end{equation}
with flip operator $\hat{A}_{mn} = \left|n\right\rangle\left\langle m\right|$. This molecular system is coupled in dipole approximation~\cite{Buhmann12a}
\begin{equation}
    \hat{H}_{\rm MF} = -\hat{\bm{d}}\cdot \hat{\bm{E}}(\rrM) =
    -\sum_{n,m} \hat{A}_{mn} {\bm d}_{mn} \cdot \hat{\bm{E}} (\rrM)\, ,
\end{equation}
where the electric field is evaluated at the molecule's position $\rrM$. In this dipole approximation, the interacting molecules are considered point molecules. For small distances (close to binding distance), typically higher orders of the multipole expansion are required~\cite{salam2009molecular}. However, we stay in the dipole approximation, discuss the fundamental results and illustrate a simple extension method towards the end of the manuscript, the discussion about finite-size effects at the end of Sec.~\ref{sec:4b}. The electromagnetic field is described by the field Hamiltonian~\cite{Buhmann12a}
 \begin{equation}
     \hat{H}_{\rm F} = \sum_{\lambda=\rm{e,m}}\int\mathrm d^3 r \int\limits_0^\infty \mathrm d \omega \, \hbar\omega\hat{\bm{f}}^\dagger_\lambda (\rr,\omega)\cdot \hat{\bm{f}}_\lambda(\rr,\omega) \, ,
 \end{equation}
with the dressed field's ladder operators $\hat{\bm{f}}_\lambda$ and $\hat{\bm{f}}^\dagger_\lambda$ containing the photonic and surface excitations (such as polarisations and polaritons), where $\lambda$ distinguishes between the electric (e) and magnetic (m) contributions. The electric field at the molecule's position will be evaluated via the Green function ${\bf{G}}(\rr,\rr',\omega)$
\begin{equation}
    \hat{\bm{E}}(\rr) = \int\limits_0^\infty \mathrm d \omega \, \sum_{\lambda=\rm{e,m}}\int\mathrm d^3 r' \, {\bf{G}}_\lambda(\rr,\rr',\omega) \cdot \hat{\bm f}(\rr',\omega) + \rm{H.c.} \,,
\end{equation}
where the Green function separates into its electric part via the projection of the total Green function ${\bf{G}}$
\begin{equation}
    {\bf{G}}_{\rm e}(\rr,\rr',\omega) = \mi \frac{\omega^2}{c^2}\sqrt{\frac{\hbar}{\pi\varepsilon_0}\IM \varepsilon(\rr',\omega) }{\bf{G}}(\rr,\rr',\omega) \, ,
\end{equation}
and magnetic part
\begin{equation}
    {\bf{G}}_{\rm m}(\rr,\rr',\omega) = \mi \frac{\omega}{c}\sqrt{\frac{\hbar}{\pi\varepsilon_0}\frac{\IM \mu(\rr',\omega)}{\left|\mu(\rr',\omega)\right|^2 } }\left[ \nabla'\times{\bf{G}}(\rr,\rr',\omega)\right]^{\rm T} \,.
\end{equation}
The Green function is the general solution of the vector Helmholtz equation~\cite{Buhmann12a}
\begin{eqnarray}
\nabla\times\frac{1}{\mu({\rr},\omega)}\nabla\times {\bf{G}}(\rr,\rr',\omega)-\frac{\omega^2}{c^2}\varepsilon(\rr,\omega)
 {\bf{G}}(\rr,\rr',\omega) = \boldsymbol{\delta}(\rr-\rr') \, , \label{eq:Helmholtz}
\end{eqnarray}
and, thus, denotes the field propagator for classical fields or the photon propagator for quantised fields. It contains information about the dielectric environment meaning the cluster in the considered case. Its particular solutions are considered directly in Sec.~\ref{sec:green}.

In the quantum-optical framework, the excitation lifetime is described via the internal molecule's dynamics, which is determined via Heisenberg's equations of motion for the flip operators~\cite{Buhmann12b}
\begin{eqnarray}
    \dot{\hat{A}}_{mn} = \frac{1}{\mi\hbar}\left[ \hat{A}_{mn},\hat{H}\right] = \mi \omega_{mn}\hat{A}_{mn}+\frac{\mi}{\hbar}\sum_k \left(\hat{A}_{mk} {\bm{d}}_{nk} - \hat{A}_{kn}{\bm{d}}_{km}\right) \cdot \hat{\bm{E}}(\rrA) \, .\label{eq:motion}
\end{eqnarray}
By solving the system of coupled equations of motion~(\ref{eq:motion}) and splitting the result into its imaginary and real parts, one obtains the atomic frequency shifts for the $n$th excited state~\cite{Ribeiro2015}
\begin{equation}
    \delta\omega_{n} = -\frac{\mu_0}{\hbar \pi} \sum_{k\ne n}\mathcal{P} \int\limits_0^\infty\mathrm d \omega \frac{\omega^2 {\bm{d}}_{nk} \cdot \IM {\bf{G}}(\rrA,\rrA,\omega)\cdot {\bm{d}}_{kn}}{\omega+\omega_{kn}} \, ,\label{eq:deltaomega}
\end{equation}
and the transition rate
\begin{equation}
    \Gamma_n = \frac{2\mu_0}{\hbar}\sum_{k<n}  \omega_{nk}^2 {\bm{d}}_{nk} \cdot \IM {\bf{G}}(\rrA,\rrA,\omega_{nk})\cdot {\bm{d}}_{kn} \, . \label{eq:Gamma}
\end{equation}
The transition rate (linewidth) $\gamma_n$ is typically detected via the excitation lifetime $\tau_n$, that are inversely related to each other $\tau_n=1/\gamma_n$. Both quantities describe the properties of the free isolated molecule. In the presence of an environment, they will change to $\tau_n\rightarrow 1/(\gamma_n +\Gamma_n)$ via the mode-coupling~(\ref{eq:Gamma}) concerning the environmental degrees of freedom. This effect is known as Purcell effect~\cite{Purcell1946}. Thus, the transition rate~(\ref{eq:Gamma}) describes the change in the free-space rate caused by the presence of the environment. At binding separation, the local mode density is typically a positive quantity. Hence, a dielectric object results in a reduction of the excitation lifetime. The simultaneous consideration of both effects is only relevant if the observed transition $\omega_{kn}$ is closed by a resonance of the dielectric object. In this case, the impact of the detuning $\delta\omega_{kn}=\delta\omega_k-\delta\omega_n$ on the local mode density $\IM \GG(\rrA,\rrA,\omega_{nk}+\delta\omega_{nk})$ gets relevant.

\subsection{Scattering Green function in bulk and near planar and spherical surfaces}\label{sec:green}
In this manuscript, we estimate the impact of the presence of a dielectric object on the superradiance between two molecules. In terms of the macroscopic quantum electrodynamics, these effects are expressed by the scattering Green functions~(\ref{eq:Green}), which include information about the electromagnetic properties of the environment. In free space, the Green function reads~\cite{Buhmann12a}
\begin{align}
\GGb(\rr,\rr',\omega) = -\frac{c^2}{3\omega^2}\boldsymbol{\delta}(\boldsymbol{\varrho})-\frac{c^2\mathrm e^{\mi\omega\varrho/c}}{4\pi\omega^2\varrho^3}\left\lbrace \left[1-\mi\frac{\omega\varrho}{c}-\left(\frac{\omega\varrho}{c}\right)^2\right]\mathbb{I}
-\left[3-3\mi\frac{\omega\varrho}{c}-\left(\frac{\omega\varrho}{c}\right)^2\right]{\bm{e}}_\varrho{\bm{e}}_\varrho\right\rbrace\,,
\end{align}
with the relative coordinate $\boldsymbol{\varrho}=\rr-\rr'=\varrho{\bm{e}}_\varrho$, its magnitude $\varrho=\left|\boldsymbol{\varrho}\right|$, its unit vector ${\bm{e}}_\varrho=\boldsymbol{\varrho}/\varrho$, and the three-dimensional unit matrix $\mathbb{I}=\operatorname{diag}(1,1,1)$, which leads to 
\begin{equation}
    \IM\,\GGb(\rr,\rr',\omega) =\frac{1}{6\pi\varrho}\sin \left(\frac{\omega\varrho}{c}\right)\mathbb{I}\,, \label{eq:free}
\end{equation}
and to the coincidence limit ($\rr'\mapsto \rr$)
\begin{equation}
    \IM\,\GGb(\rr,\rr,\omega) = \frac{\omega}{6\pi c}\mathbb{I}\,. \label{eq:freecoin}
\end{equation}
By inserting this result into the spectral detuning~(\ref{eq:deltaomega}), one obtains the Lamb shift~\cite{Lamb1947}, and into the change of the transition rate~(\ref{eq:Gamma}), one obtains the well-known Einstein coefficient.~\cite{Dung2003,messiah2014quantum} These impacts of the quantum vacuum are included via a renormalised response function.

\subsubsection{Fresnel scattering at planar surfaces}
For large cluster radii, the electromagnetic scattering at the cluster can be approximated by the Fresnel reflection at a planar interface. 
In this limit, the scattering Green function is given by~\cite{Buhmann12a}
\begin{eqnarray}
\GGp(\rr,\rr',\omega) = \frac{\mi}{8\pi^2}\int\frac{\md^2 k^\parallel}{k_1^{\perp}}\me^{\mi \kk^\parallel\cdot (\rr-\rr')+\mi k_1^\perp (z+z')}
\left[ r_s \ee_{s+}^1\ee_{s-}^1 +r_p \ee_{p+}^1\ee_{p-}^1\right] \,, \label{eq:GreenPlanar}
\end{eqnarray}
with the Fresnel reflection coefficients
\begin{eqnarray}
r_s = \frac{k_1^\perp -k_2^\perp}{k_1^\perp+k_2^\perp}\,,\quad
r_p = \frac{\varepsilon_2 k_1^\perp -\varepsilon_1 k_2^\perp}{\varepsilon_2 k_1^\perp+\varepsilon_1 k_2^\perp}\,,
\end{eqnarray}
the wave vector parallel to the plane $\kk^\parallel \perp \ee_z$ and its component towards $z$ direction
$
k^\perp_j = \sqrt{\varepsilon_j  \omega^2/c^2-{k^\parallel}^2}
$. For the considered geometry, the Green tensor~(\ref{eq:Greenplanar}) simplifies in the non-retarded limit at equal positions to
\begin{eqnarray}
\IM\GGp(\rr,\rr,\omega)=\frac{c^2}{32\pi\omega^2z^3}\IM r_p  \begin{pmatrix}1 & 0 & 0\\
0 & 1  & 0 \\
0 & 0 & 2 
\end{pmatrix}\,,\label{eq:GP}
\end{eqnarray}
with the distance between the surface and the molecule $z$. In the non-retarded limit, the reflection coefficient for $s$-polarised waves vanishes $r_s=0$, and the reflection coefficient for $p$-polarised waves simplifies to
\begin{equation}
    r_p \approx \frac{\varepsilon_2 -\varepsilon_1}{\varepsilon_2 +\varepsilon_1 }\,,
\end{equation}
which is, thus, $k$ independent. Details on its derivation can be found in App.~\ref{app:plane}. For the consideration of the superradiance, only the $zz$-component is required, which reads in the non-retarded limit
\begin{eqnarray}
  G_{zz}(\rr,\rr',\omega) = \frac{1}{4\pi}\frac{r_p c^2}{\omega^2}\int \md k^\parallel \me^{-2 k^\parallel z} {k^\parallel}^2 J_0(k^\parallel x)\,.
\end{eqnarray}
Finally,  we find
\begin{eqnarray}
    \IM G_{zz}(\rr,\rr',\omega) =-\frac{1}{4\pi k_0^2z^3}
\IM r_p \frac{(x/z)^2-8}{\left((x/z)^2+4\right)^{5/2}}
\,, \label{eq:planar}
\end{eqnarray}
where we assumed both molecules to be located at the same distance from the surface $z$ and an in-plane separation $x$.
\subsubsection{Mie scattering at spherical surfaces}
As mentioned above, we consider a pentacene molecule in front of an Argon nano-droplet. 
This droplet is modelled as a dielectric sphere. This assumption yields the requirement of the Mie scattering to obtain the impact of the curvature onto the decay rate~(\ref{eq:Gamma}). 
In this case, the scattering Green function for source and final points outside the sphere of radius $R$ and permittivity  $\varepsilon(\omega)$ reads~\cite{339756,Scheel2008}
\begin{eqnarray}
    &&\GGs(\rr,\rr',\omega) = \frac{\mi k_0}{4\pi} \sum_{p=\rm{e,o}} \sum_{l=1}^\infty \sum_{m=0}^l (2-\delta_{m0})\frac{2l+1}{l(l+1)}\frac{(l-m)!}{(l+m)!} \nonumber\\
    &&\times \left[ r_s {\bm{M}}_{pml}(k_0) {\bm{M}}_{pml}'(k_0) + r_p {\bm{N}}_{pml}(k_0) {\bm{N}}_{pml}'(k_0)\right] \, ,\label{eq:Green}
\end{eqnarray}
where the prime sign denotes that the primed argument $\rr'$ has to be used in the vector wave functions. The reflection coefficients for $s$-polarised waves reads as
\begin{equation}
    r_s = -\frac{k \eta_l(k R)j_l(k_0R)-k_0\eta_l(k_0R)j_l(kR)}{k \eta_l(k R) h_l^{(1)}(k_0R)-k_0 \zeta_l(k_0 R)j_l(kR)} \,,\label{eq:rs}
\end{equation}
and for $p$-polarised waves as
\begin{equation}
    r_p = -\frac{k \eta_l(k_0R)j_l(kR)-k_0\eta_l(kR)j_l(k_0R)}{k \zeta_l(k_0R)j_l(kR)-k_0\eta_l(kR)h_l^{(1)}(k_0R)} \,, \label{eq:rp}
\end{equation}
with the spherical Bessel and Hankel function of the first kind $j_l(x)$ and $h_l^{(1)}(x)$, respectively, and the Ricatti functions
\begin{equation}
    \eta_l(x) = \frac{1}{x}\frac{\mathrm d \left[xj_l(x)\right]}{\mathrm d x}\,,\qquad \zeta_l(x) = \frac{1}{x}\frac{\mathrm d \left[xh_l^{(1)}(x)\right]}{\mathrm d x}\,.
\end{equation}
Furthermore, the scattering Green's function~(\ref{eq:Green}) requires the spherical vector wave functions~\cite{339756}
\begin{eqnarray}
\lefteqn{{\bm{M}}_{{\rm{{}_o^e}}ml}(k) = \mp\frac{m}{\sin\vartheta}h_l^{(1)}(kr)P_l^m(\cos\vartheta)\begin{matrix}
\sin\\\cos\end{matrix}m\varphi {\bm{e}}_\vartheta-h_l^{(1)}(kr)\frac{\mathrm d P_l^m(\cos\vartheta)}{\mathrm d\vartheta}\begin{matrix}\cos\\\sin\end{matrix} m\varphi {\bm{e}}_\varphi \,,}\label{eq:M}\\ 
\lefteqn{{\bm{N}}_{{\rm{{}_o^e}}ml}(k) = \frac{l(l+1)}{kr} h_l^{(1)}(kr)P_l^m(\cos\vartheta)\begin{matrix} \cos\\\sin\end{matrix} m\varphi {\bm{e}}_r}\nonumber\\
&&+\frac{1}{kr}\frac{\mathrm d r h_l^{(1)}(kr)}{\mathrm d r} \left[\frac{\mathrm d P_l^m(\cos\vartheta)}{\mathrm d \vartheta}\begin{matrix}\cos\\\sin\end{matrix}m\varphi {\bm{e}}_\vartheta \mp \frac{mP_l^m(\cos\vartheta)}{\sin\vartheta}\begin{matrix}\sin\\\cos\end{matrix}m\varphi {\bm{e}}_\varphi\right]\,. \label{eq:N}
\end{eqnarray}
with the vacuum wave number $k_0=\omega/c$ and the wave number inside the sphere $k=k_0\sqrt{\varepsilon(\omega)}$. The vectors ${\bm{e}}_r$, ${\bm{e}}_\vartheta$ and ${\bm{e}}_\varphi$ are the mutually orthogonal unit vectors. 
The reflection coefficients~(\ref{eq:rs}) and (\ref{eq:rp}) depend neither on the orientation $\rm{e,o}$ nor on $m$. 
This allows us to derive the corresponding sums together with the addition theorem for Legendre polynomials~\cite{gradshteyn2007,Buhmann2004}
\begin{eqnarray}
    \lefteqn{P_l(\cos\varphi_1\cos\varphi_2 + \sin\varphi_1\sin\varphi_2\cos\Theta)=}\nonumber\\
    &&\sum_{m=0}^l (2-\delta_{m0})\frac{(l-m)!}{(l+m)!} P_m^l(\cos\varphi_1)P_m^l(\cos\varphi_2)\cos \left(m \Theta\right)\, , \label{eq:additiontheorem}
\end{eqnarray}
which reduces the Green function for Mie scattering in the coincidence limit ($\rr'\mapsto\rr$) to
\begin{eqnarray}
\lefteqn{\GGs(\rr,\rr,\omega) = \frac{\mi k_0}{8\pi}  \sum_{l=1}^\infty(2l+1)  \left[ r_s \left[h_l^{(1)}(k_0r)\right]^2 \left({\bm{e}}_\vartheta {\bm{e}}_{\vartheta}+{\bm{e}}_\varphi{\bm{e}}_\varphi\right)\right.}\nonumber\\
 &&\left.+ r_p \Biggl\lbrace 2\frac{l(l+1)}{k_0^2r^2}\left[h_l^{(1)}(k_0r)\right]^2 {\bm{e}}_r{\bm{e}}_{r} 
 + \left[\frac{1}{k_0r}\frac{\mathrm d r h_l^{(1)}(k_0r)}{\mathrm d r}\right]^2\left({\bm{e}}_\vartheta {\bm{e}}_{\vartheta}+{\bm{e}}_\varphi{\bm{e}}_\varphi\right) \Biggr\rbrace\right] \, .\label{eq:Greenr=r'}
\end{eqnarray}
In the non-retarded limit ($\left|\rr\right| -R \ll R$), the reflection coefficients~(\ref{eq:rs}) and (\ref{eq:rp}) can be approximated by
\begin{align}
    r_s &\approx 0 \,,\\
    r_p &\approx \mi \frac{l+1}{(2l+1)!!(2l-1)!!}\frac{\varepsilon(\omega)-1}{l\varepsilon(\omega)+1+l} \left(\frac{\omega R}{c}\right)^{2l+1}\,,
\end{align}
with the double factorial $(2n+1)!! = \prod_{k=1}^n (2k+1)$, by using the asymptotic forms of the spherical Bessel and first kind Hankel function~\cite{abramowitz+stegun}
\begin{align}
j_l(z) \simeq \frac{z^l}{(2l+1)!!}\,,\quad
h_l^{(1)}(z) \simeq \frac{-\mi (2l-1)!!}{z^{l+1}}\,,
\end{align}
respectively, for $\left|z\right|\ll 1$. Finally, the scattering dyadic Green function in the non-retarded limit reads
\begin{align}
    \GGs(\rr,\rr,\omega) = \frac{c^2}{8\pi \omega^2 r^3}  \sum_{l=1}^\infty l(l+1)\frac{\varepsilon-1}{l\varepsilon+l+1}\left(\frac{R}{r}\right)^{2l+1}\left[ 2 (l+1) {\bm{e}}_r{\bm{e}}_{r} +l
 \left({\bm{e}}_\vartheta {\bm{e}}_{\vartheta}+{\bm{e}}_\varphi{\bm{e}}_\varphi\right)\right] \, ,\label{eq:Greenfinal}
\end{align}
which agrees with the Green function for the Casimir--Polder potential for an atom close to a sphere. Details on this calculation are given in App.~\ref{app:sphere}.

For large sphere radii $R$, the main contribution of the $l$-sum of the Green function~(\ref{eq:Greenfinal}) come 
from large $l$ values and the scattering Green function reduces to a planar surface~\cite{Buhmann12a,Tomas1995}
\begin{equation}
 \GGp (\rrA,\rrA,\omega) = \frac{c^2}{32\pi\omega^2z^3} \frac{\varepsilon(\omega)-1}{\varepsilon(\omega)+1} \operatorname{diag}(1,1,2) \, ,\label{eq:Greenplanar}
\end{equation}
consisting of the Fresnel reflection coefficient at the surface $(\varepsilon-1)/(\varepsilon+1)$ and the diagonal matrix $\operatorname{diag}(1,1,2)$ denoting the anisotropy of the space. Here, the interface is located in the $x$-$y$-plane with the distance between the molecule and the surface $z=r-R$.

The $rr$-component in the non-retarded limit of two-point scattering Green function required for computing the superradiance is given by
\begin{eqnarray}
\IM\GG_{rr}(\rr,\rr',\omega) = \frac{1}{4\pi k_0^2r^3} \sum_{l=1}^\infty l(l+1)^2\IM\left[\frac{\varepsilon(\omega)-1}{l\varepsilon(\omega)+1+l}\right]  \left(\frac{R}{r}\right)^{2l+1} P_l\left(\cos\vartheta'\right) \,.\label{eq:Grrp}
\end{eqnarray}

\begin{figure}[t]
    \centering
    \begin{subfigure}[t]{0.49\columnwidth}
        \centering
        \includegraphics[width=0.99\columnwidth]{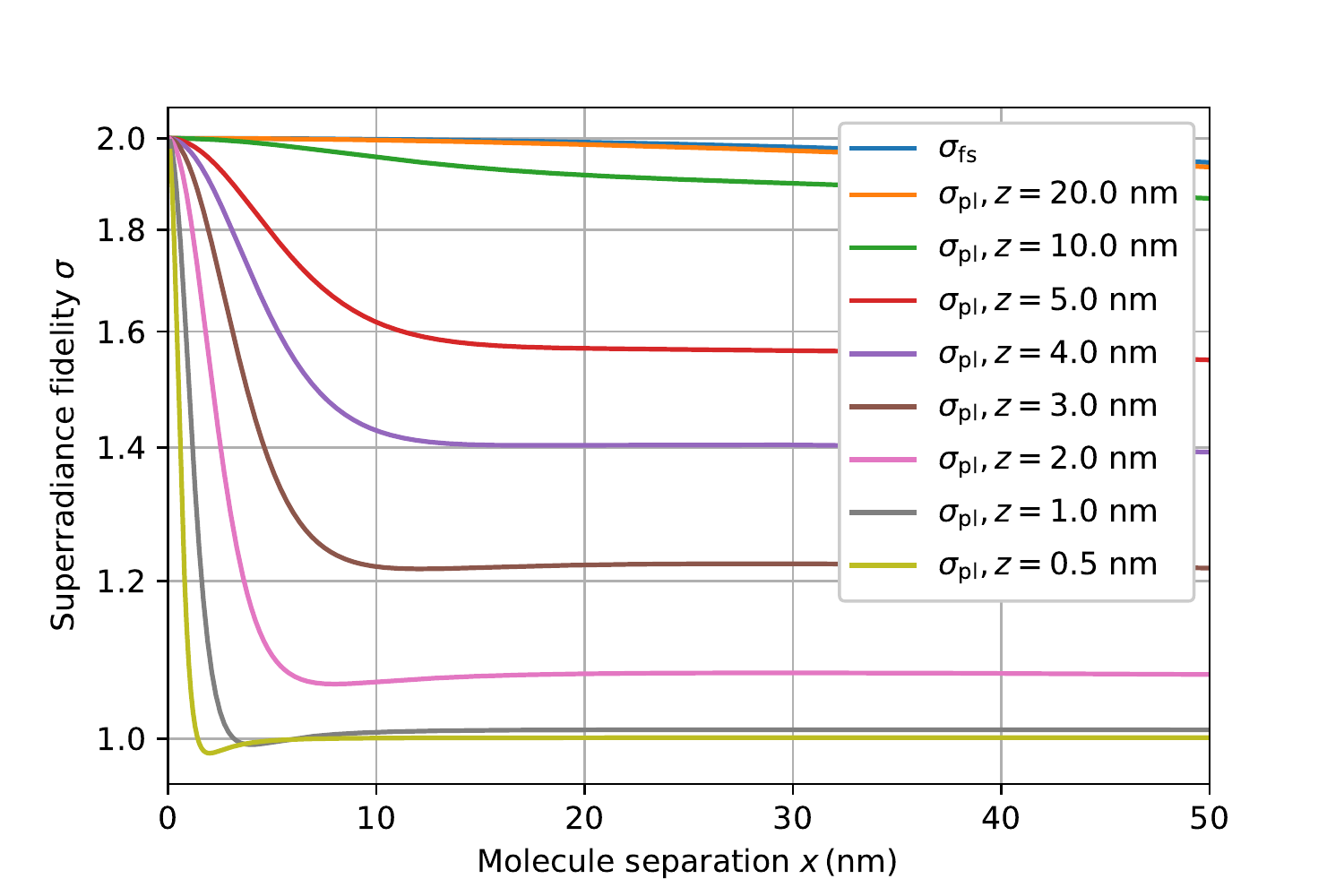} 
        \caption{On Argon} 
    \end{subfigure}
    \hfill
    \begin{subfigure}[t]{0.49\columnwidth}
        \centering
        \includegraphics[width=0.99\columnwidth]{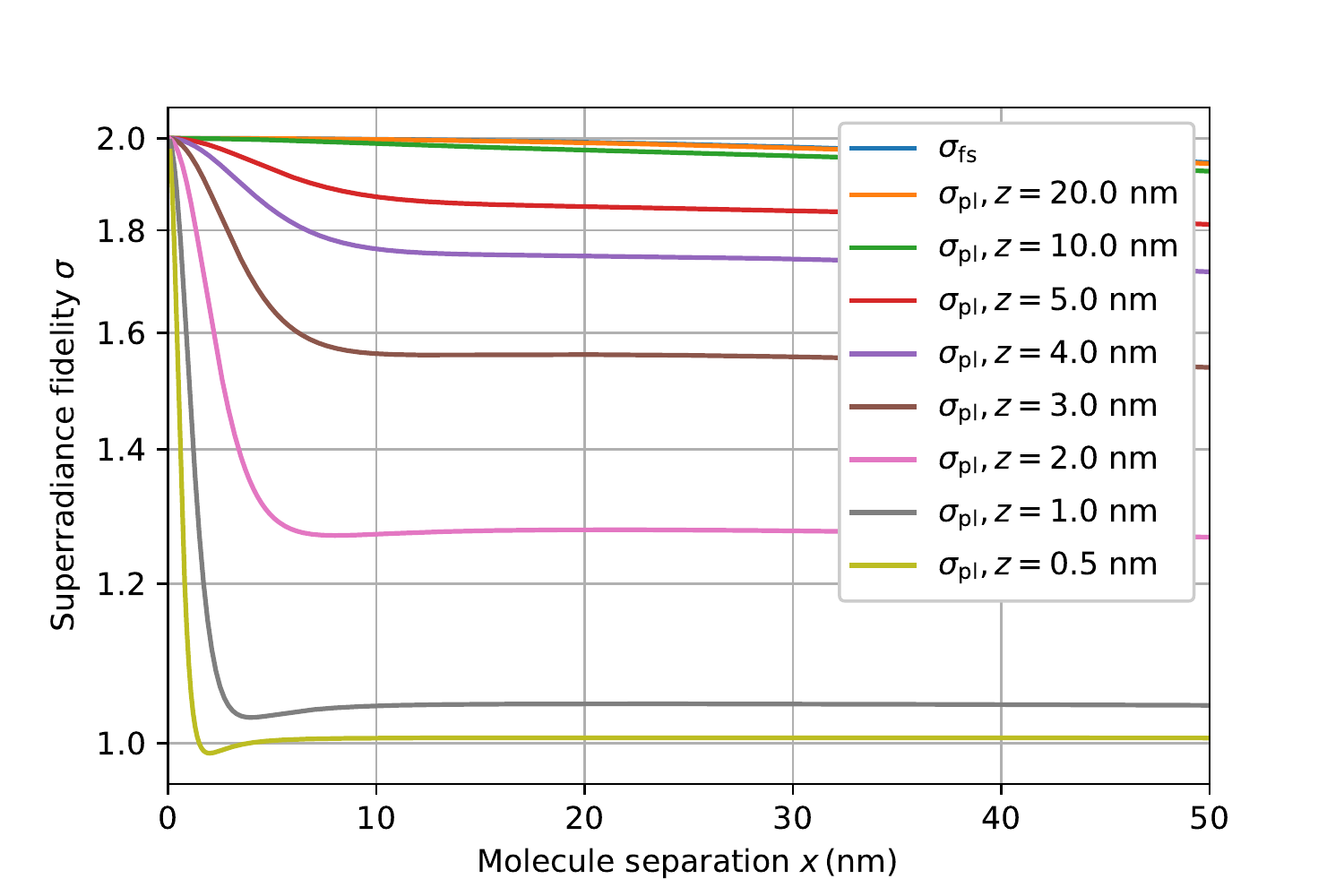} 
        \caption{On Neon}
    \end{subfigure}
    \caption{Superradiance fidelity $\sigma$ for two molecules separated by a distance $x$ for different distances to a surface $z$ at 20~nm (orange line), 10~nm (green line), 5~nm (red line), 4~nm (purple line), 3~nm (brown line), 2~nm (pink line), 1~nm (grey line) and at roughly binding separation 0.5~nm (olive line). For comparison, the free-space fidelity $\sigma_{\rm fs}$ is also shown. In particular, we have chosen an Argon cluster (a) and a Neon cluster (b), and a radiation frequency $\omega=3.4753\cdot 10^{15}\,\rm{ rad/s}$ which corresponding to $S_1\rightarrow S_0$-transition of pentacene, similar to Ref.~\cite{doi:10.1063/1.5097553}}
    \label{fig:sr_plate}
\end{figure}

\section{Superradiance of two point-like molecules near an interface}

The superradiance is characterised by the relation~(\ref{eq:sigma}). In this description, the superradiance rate is almost independent of the considered molecules. Due to the rotational average, see App.~\ref{app:rot}, the dipole transition factorises, and only the frequency of the radiation field remains. In the considered scenarios, with symmetry [$\GG(\rr,\rr')=\GG(\rr',\rr)$], the superradiance rate in free-space (subscribed fs) can be described as
\begin{equation}
    \sigma_{\rm fs} = 1+\frac{\IM\operatorname{Tr}\GGb(\rr,\rr',\omega)}{\IM\operatorname{Tr}\GGb(\rr,\rr,\omega)}=1+ \frac{\sin k_0x}{k_0x}\,,\label{eq:sigmafs}
\end{equation}
with the non-retarded free-space Green function~(\ref{eq:free}), its coincidence limit~(\ref{eq:freecoin}), the distance between the molecules $x$ and the wave vector of the considered transition $k_0=\omega/c$. It shows that the relevant length scale for separating the molecules is the wavelength of the radiation. Thus, superradiance between two molecules in free space is a long-range phenomenon governed by the respective transition's wavelength. Superradiance is suppressed for particle separations longer than the wavelength, and the particles will react individually. Superradiance only occurs for separations below the wavelength.

\begin{figure}[t]
    \centering
    \begin{subfigure}[t]{0.49\columnwidth}
        \centering
        \includegraphics[width=0.99\columnwidth]{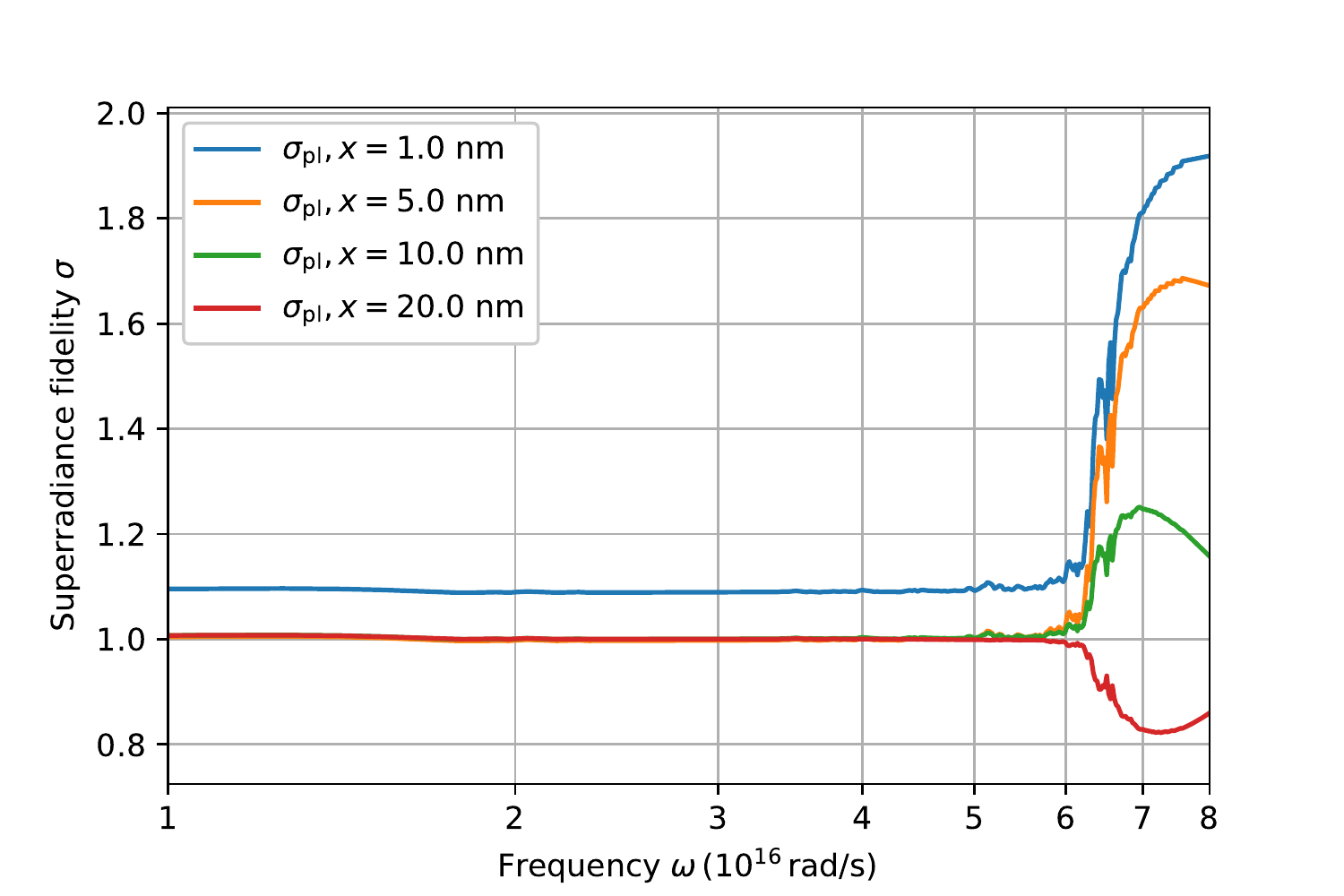} 
        \caption{On Argon} 
    \end{subfigure}
    \hfill
    \begin{subfigure}[t]{0.49\columnwidth}
        \centering
        \includegraphics[width=0.99\columnwidth]{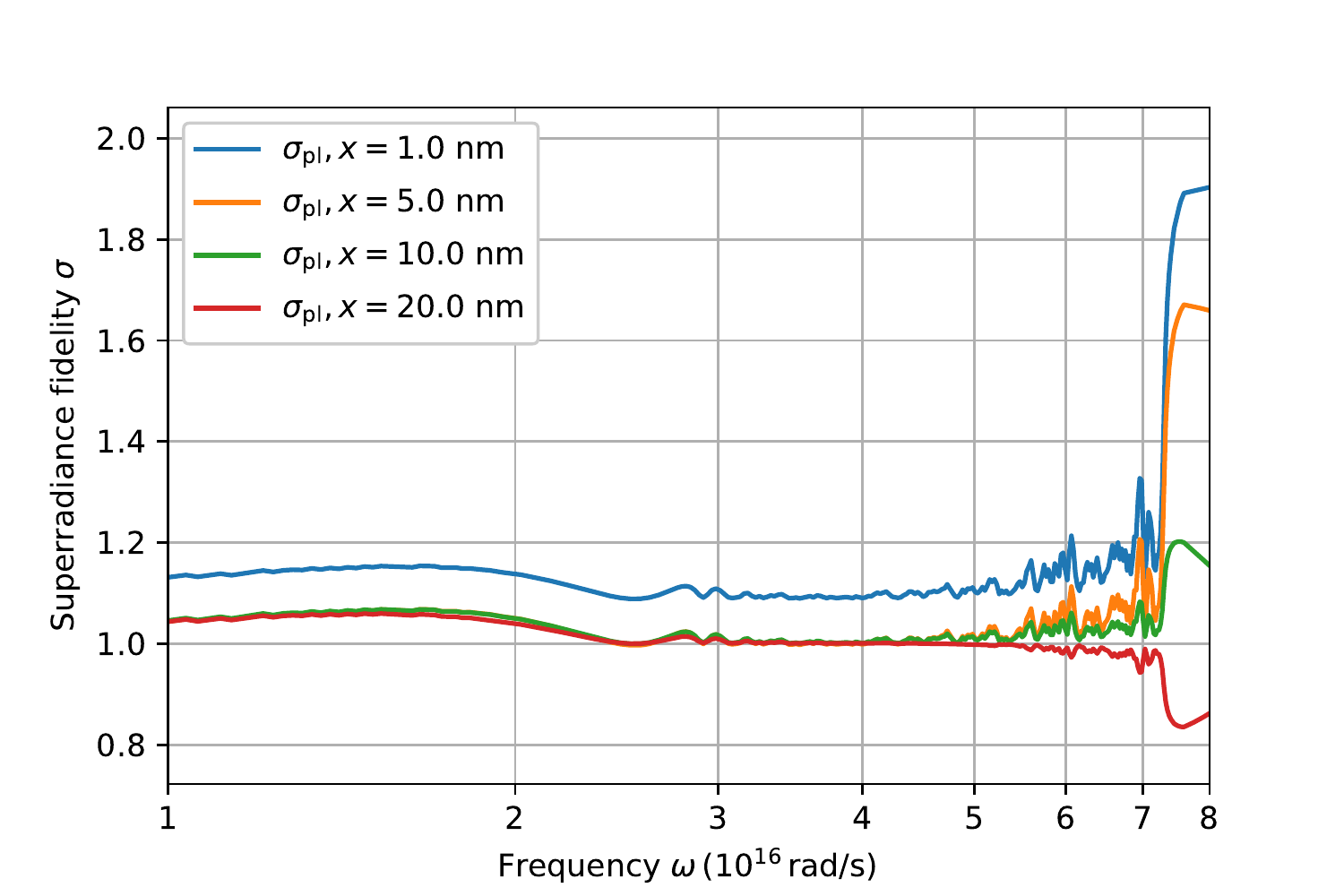} 
        \caption{On Neon}
    \end{subfigure}
    \caption{Superradiance fidelity $\sigma$ for two molecules separated by a distance $x$ at binding distance to the surface $z=0.5~\rm{nm}$ depending on the radiation frequency determined via the dielectric response of Argon (a) and Neon (b).}
    \label{fig:sr_platefreq}
\end{figure}
\subsection{Superradiance near planar interfaces}\label{sec:plate}

The superradiance fidelity in front of a plate (subscribed pl)
is found to be
\begin{eqnarray}
 \sigma_{\rm pl} &=& 1+ \frac{\IM\GGb_{zz}(\rr,\rr')+\IM\GG_{{\rm pl},zz}(\rr,\rr')}{\IM\GGb_{zz}(\rr,\rr)+\IM\GG_{{\rm pl},zz}(\rr,\rr)}\nonumber\\
 &=&1+\frac{96z^3 \left[\frac{4}{3}k_0^2\sin\left(k_0x\right) \sqrt{x^2+4z^2}\left(z^2+x^2/4\right)^2+x\left(z^2-x^2/8\right)\IM r_p\right]}{x\left(8k_0^3 z^3+ 3\IM r_p\right)\left(x^2+4z^2\right)^{5/2}}\nonumber\\
 &=&1+\frac{128 \lambda^3\left[ k_0^2x^2\sin k_0x\left(\lambda^2+1/4\right)^2 \sqrt{4\lambda^2+1}+\frac{3}{32}\IM r_p \left(8\lambda^2-1\right)\right]}{\left(8k_0^3x^3\lambda^3+3\IM r_p\right)\left(4\lambda^2+1\right)^{5/2}}\,,
\end{eqnarray}
with the non-retarded free-space Green function~(\ref{eq:free}), its coincidence limit~(\ref{eq:freecoin}), the non-retarded Green function for the planar interface~(\ref{eq:planar}), its coincidence limit~(\ref{eq:GP}), the ratio between separation of both molecules and their distance to the surface $\lambda=z/x$. The consideration of the non-retarded Green functions restricts the model to non-radiative decays~\cite{4683624,doi:https://doi.org/10.1002/9780470142561.ch1}, which are dominant for the considered scenario of weakly responding substrates and small distances between the particles and the surface. For small values of this ratio, which corresponds to the case where the molecules are closer to the surface than separated from each other, the result can be expanded in a Taylor series leading to
\begin{eqnarray}
    \sigma_{\rm pl} = 1+ \left(\frac{8}{3\IM r_p}k_0^2x^2\sin k_0x -4\IM r_p\right)\lambda^3 
    \approx 1+ \left(\sigma_{\rm fs}-1\right) \frac{8 k_0^3 z^3}{3 \IM r_p} \,.\label{eq:sigmapl}
\end{eqnarray}
In the last step, we have inserted the free-space fidelity~(\ref{eq:sigmafs}) and used $\lambda \ll1$. It shows that the presence of the surface causes the free-space fidelity to be strongly suppressed. The relevant length scale can be identified as the distance to the surface.

\begin{figure}
    \centering
    \includegraphics[width=0.8\columnwidth]{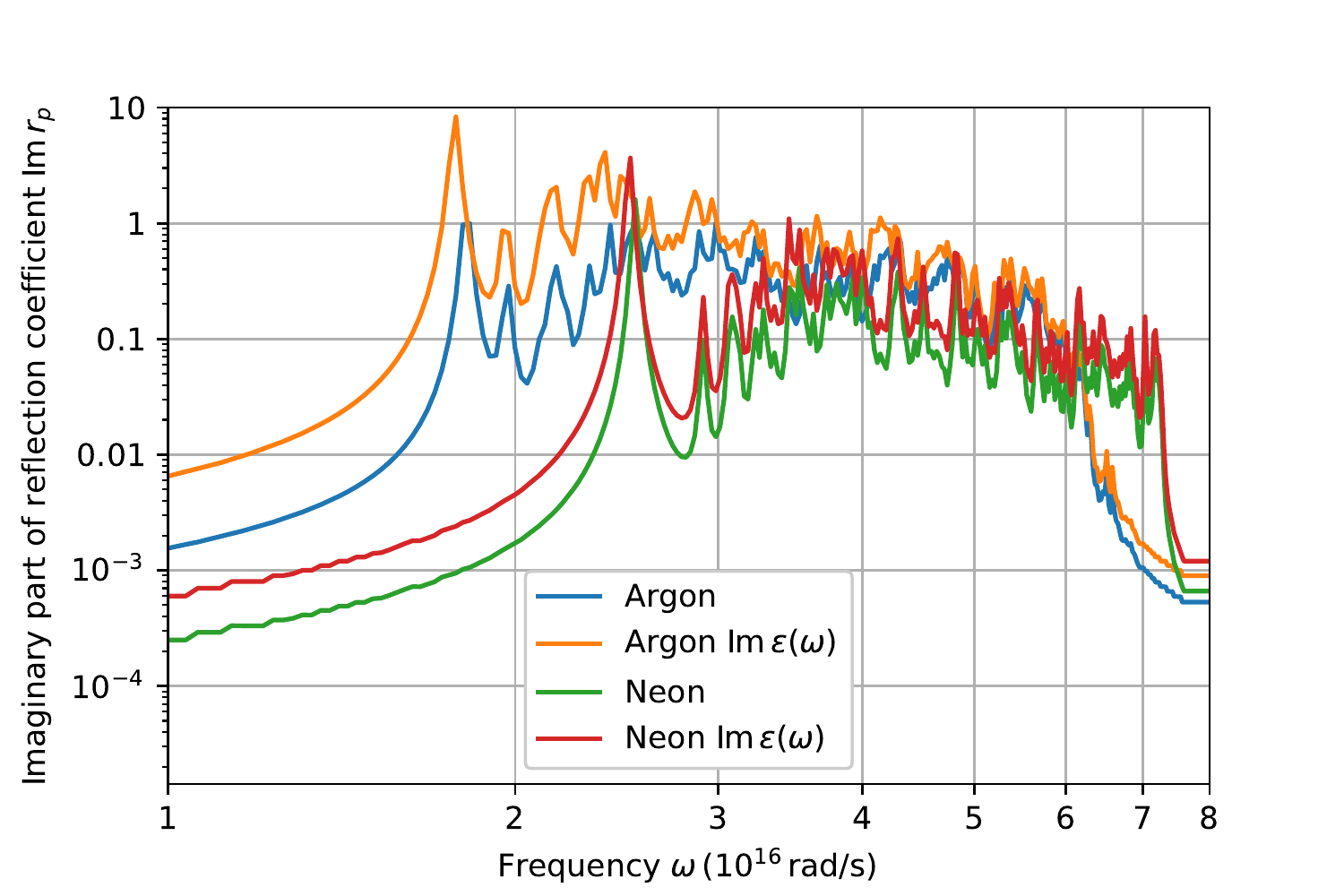}
    \caption{Imaginary part of the reflection coefficients for Argon (blue line) and Neon (green line). For comparison, the imaginary parts of corresponding dielectric functions for Argon (orange line) and Neon (red line) are plotted.}
    \label{fig:reflection}
\end{figure}

Figure~\ref{fig:sr_plate} illustrates the observed behaviour of the superradiance fidelity for two pentacene molecules attached to an Argon (a) and a Neon surface (b) for the $S_1\rightarrow S_0$-transition with a radiation frequency $\omega=3.4753\cdot 10^{15}\,\rm{ rad/s}$. One observes that the relevant length scale is the binding distances to the surface. As this length is very short, it strongly reduces the possibility of observing superradiance compared to the free-space case. Further, one sees the impact of the surface drops for larger surface--molecule distances caused by the $z^{-3}$-power law of the surface potential~\cite{Scheel2008} contributing to the superradiance fidelity. Because of the low dielectric response of the considered materials (Argon and Neon), a molecule--surface separation of merely 20~nm restores the free-space superradiance fidelity. Thus, for larger separations, the impact of the surface can be neglected. The observed reduction of the superradiance fidelity with increasing molecule--molecule distance is similar for all frequencies. However, the dependence on the imaginary part of the reflection, which is related to surface plasmons, causes the superradiance fidelity to depend strongly on the dielectric response of the supporting material.  The observed superradiance fidelity plateaus illustrate the free-space rate's functional dominance. The drop-down to 1 occurs for separations larger than plotted. It can be observed that Eq.~(\ref{eq:sigmapl}) satisfies the expected limit $\lim_{x\mapsto\infty} \sigma_{\rm pl} = 1$. 
Figure~\ref{fig:sr_platefreq} illustrates the spectral dependence of the superradiance fidelity for two molecules at different distances between both (1, 5, 10 and 20~nm).
The free-space contribution can be observed to dominate the superradiance fidelity at low frequencies, and at higher frequencies, the material properties start playing a role. Interestingly, the impact of the material response is strong for more weakly responding materials, which can be seen by comparing Fig.~\ref{fig:sr_platefreq} (a) and (b) together with the imaginary part of the reflection coefficient depicted in Fig.~\ref{fig:reflection}. This behaviour illustrates the dominant role of the optical mode density [coincidence limit ${\bf{G}}(\rr,\rr)$], contributing inversely to the superradiance fidelity.
The observed behaviours far beyond the peak have to be interpreted cautiously, as the GW-BSE calculations were not carefully converged for very high energies. Remarkably, for even larger molecule--molecule distances, the superradiance fidelity flips sign and reaches a minimum at a 20\% reduced radiation fidelity compared to the single-molecule in free space.  
The Green function implies two possibilities for changing the sign and leading to a reduction of the superradiance: (i) by going to separations larger than the wavelength, where the superradiance fidelity will follow the damped wave propagation, as it is the case in free space, see Eq.~(\ref{eq:sigmafs}), or (ii) by special geometric arrangements of the dipoles and their corresponding images. The latter describes the origin of the effect observed here. However, the exact conditions are hard to illustrate explicitly due to the curvature effects of the interface.  Other effects for sign changes of the superradiance fidelity, such as position-dependent spectral shifts and non-radiative decoherence channels, exist beyond the Green function manipulation.

\begin{figure*}[t]
    \centering
        \begin{subfigure}[t]{0.49\columnwidth}
        \centering
        \includegraphics[width=0.99\columnwidth]{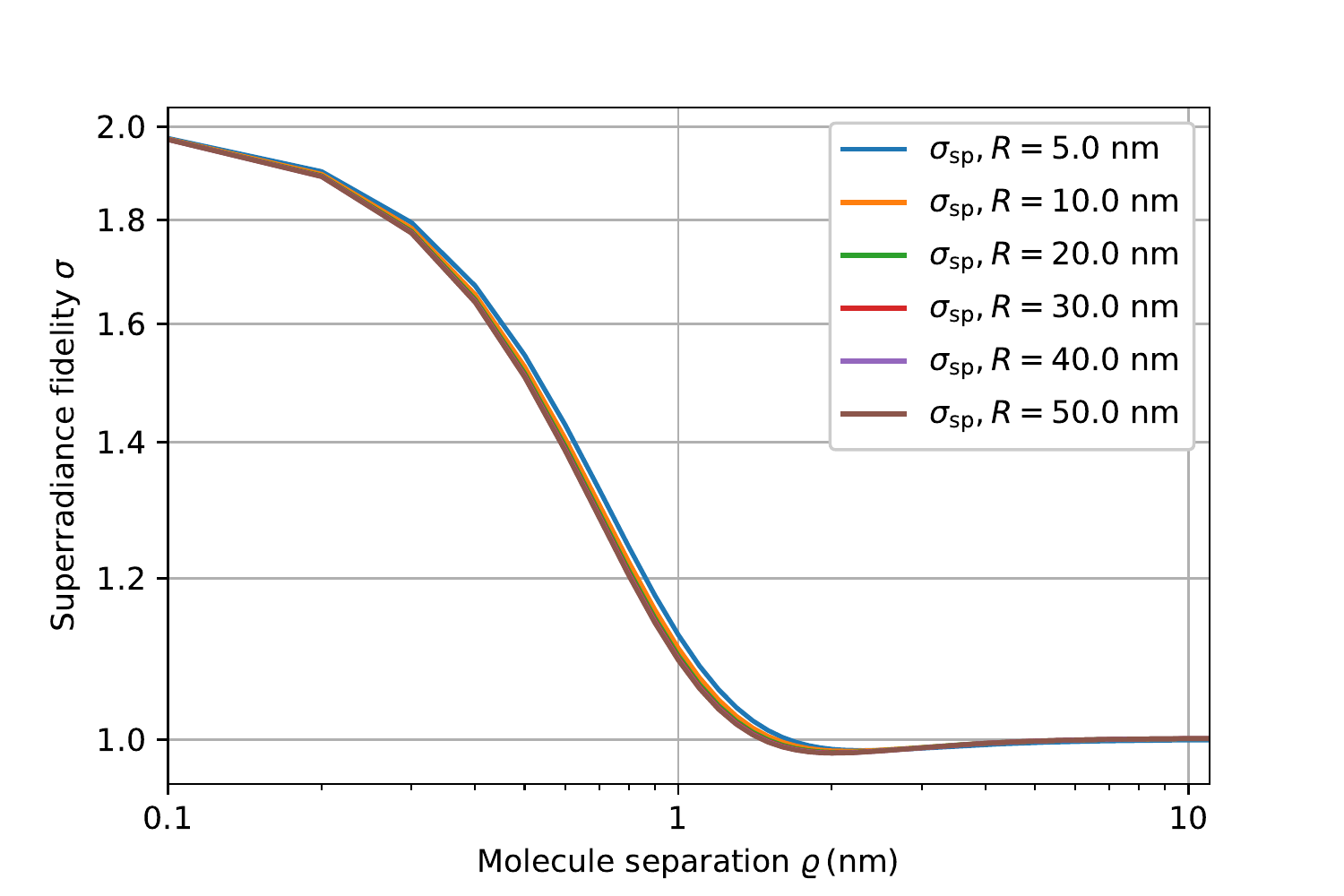} 
        \caption{Spatial separation} 
    \end{subfigure}
    \hfill
    \begin{subfigure}[t]{0.49\columnwidth}
        \centering
        \includegraphics[width=0.99\columnwidth]{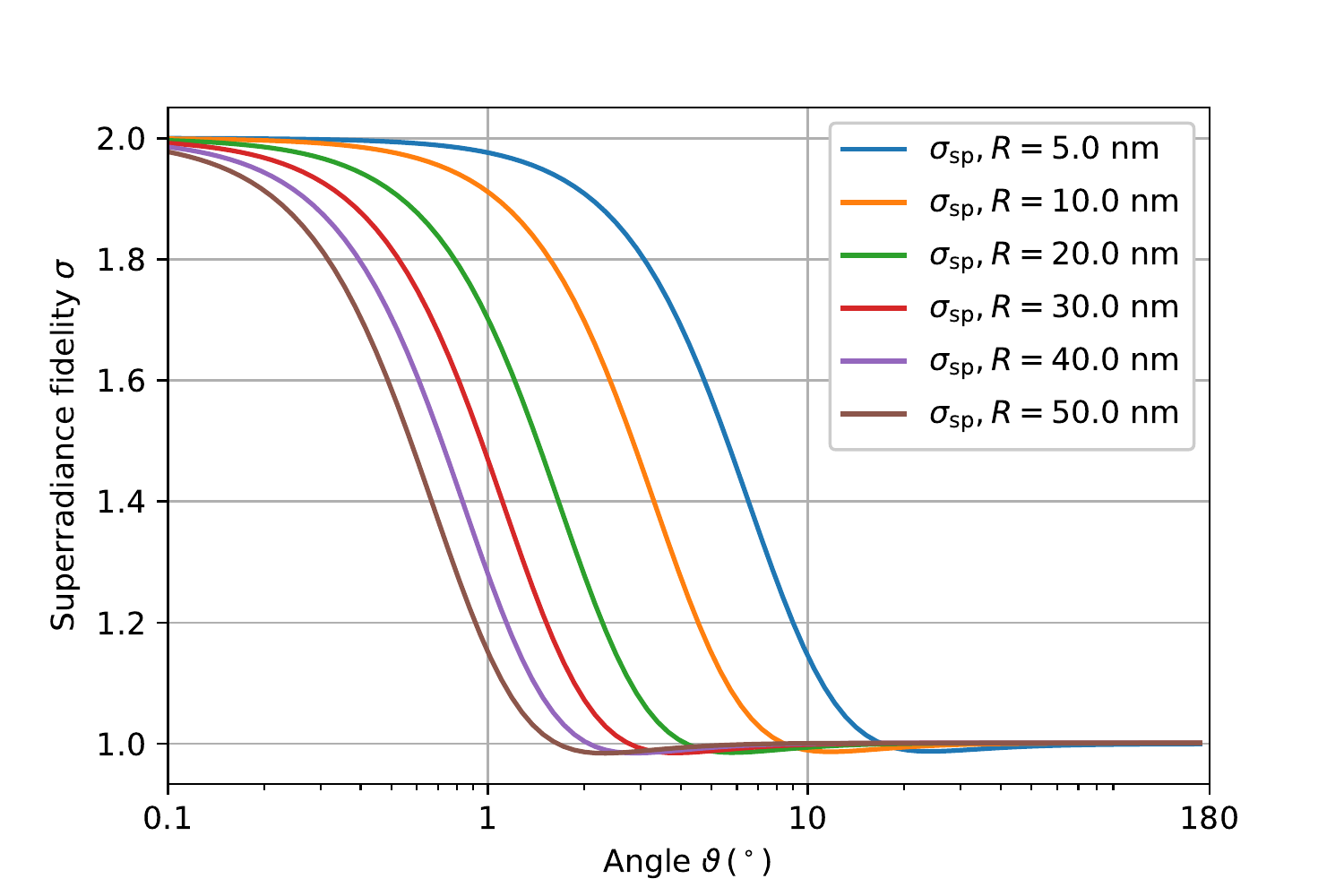} 
        \caption{Angular separation}
    \end{subfigure}
    \caption{Superradiance fidelity $\sigma_{\rm sp}$ for two pentacene molecules attached at 0.5~nm separation to a spherical Argon cluster with different curvature radii $R$: 5~nm (blue curve), 10~nm (orange), 20~nm (green), 30~nm (red), 40~nm (purple) and 50~nm (brown) plotted against the separation between molecules (a) and on the angle between both molecules and the centre of the cluster (b).}
    \label{fig:sr_sphere}
\end{figure*}

\subsection{Superradiance near spheres}\label{sec:4b}
Finally, we analyse the impact of the curvature on the superradiance. To describe this case, we insert the results of the non-retarded spherical Green function~(\ref{eq:Greenfinal}) and (\ref{eq:Grrp}) into the superradiance fidelity~(\ref{eq:sigma}) leading to the expression (subscript sp)
\begin{eqnarray}
 \sigma_{\rm sp} &=& 1+ \frac{\IM\GGb_{zz}(\rr,\rr')+\IM\GG_{{\rm sp},rr}(\rr,\rr')}{\IM\GGb_{zz}(\rr,\rr)+\IM\GG_{{\rm sp},rr}(\rr,\rr)}\,,
\end{eqnarray}
the non-retarded free-space Green function~(\ref{eq:free}), its coincidence limit~(\ref{eq:freecoin}), the non-retarded Green function for the spherical interface~(\ref{eq:Grrp}) and its coincidence limit~(\ref{eq:Greenfinal}). In analogy to the planar system, introduced in Sec.~\ref{sec:plate}, the consideration of the non retarded Green functions restricts the model to non-radiative decays~\cite{4683624,doi:https://doi.org/10.1002/9780470142561.ch1}, which are dominant for the considered scenario of weakly responding substrates and small distances between the particles and the surface.
Due to the summation over the spherical modes, a compact result is not obtainable. Instead, we numerically analyse the dependence of the superradiance fidelity for two pentacene molecules bounded to an Argon cluster with different radii $R$. The results are depicted in Fig.~\ref{fig:sr_sphere}, where one observes that the small clusters with radii below 10 nm slightly reduce the superradiance fidelity, which is, in our opinion, not observable because it has a relative impact below 1\%. Figure~\ref{fig:sr_sphere} (b) shows the superradiance rate depending on the angle $\vartheta$ between both molecules and the centre of the sphere according to the spherical Green function~(\ref{eq:Green}), (a) depicts the same physical situation but along the arc length between both molecules. Based on both figures, it can be observed that: (i) the superradiance fidelity is almost independent of the cluster diameter, and (ii) for relevant molecule separations (larger than 1~nm), the superradiance is entirely suppressed. The system can effectively be treated as a plane surface for larger curvature radii. One sees that the curvature of the surface is unimportant for the superradiance, and only the molecule--molecule distance plays a significant role. The observed effect is caused by the small separation between the molecules and the surface, see Sec.~\ref{sec:plate}.

To include finite-size effects, the electromagnetic scattering process can be smeared out over the whole molecule by a spatially dependent polarisability~\cite{Parsons2009} distribution and averaging over the weighted pairwise interactions~\cite{CPextendedJF,Brand15,D0CP02863K,doi:10.1021/acs.jpca.8b01989}. This approach can be applied to the considered system due to the spatial delocalisation of excitations in pentacene. However, the result will be equivalent to the obtained fidelities, but the centre-to-centre distance has to be exchanged with the shortest distance between both molecules~\cite{D0CP02863K}.

\section{Conclusions}
We have illustrated a theoretical approach to separate a complex system into its components and calculated a theory to estimate the change of excitation lifetimes in the presence of absorbing media. Furthermore, we derived an effective model to predict the superradiance fidelity of two molecules depending on their geometric arrangement concerning their separation and distance to the surface. In addition, we illustrated a possible extension of the model to include finite-size and orientational effects.

The derived model has been applied to the superradiance between two molecules (in particular, the optical transition of pentacene) attached to an Argon and a Neon cluster and analysed the superradiance fidelity depending on the molecule separation and their distance to the surface. We found a strong suppression for the superradiance of molecules bound to a surface and that an enhancement can only occur for densely packed molecules. Furthermore, we analysed the impact of surface curvature on the superradiance, where we did not find any remarkable influence caused by the small distance between molecules and the surface. The introduced approach is not restricted to noble gas surfaces, as long as there is no strong electronic wave-function overlap between molecule and surface. Formally, every surface can be described by inserting the corresponding dielectric function where surface plasmons feature explicitly as resonances in the reflection coefficients. In addition, this approach can be directly applied to atoms or small clusters.

The introduced approach~(\ref{eq:sigma}) can be extended directly to few- or many-molecule systems sharing the excitation due to the additivity of the local-mode density. However, the numbers of terms in the numerator grow quadratically with the number of interacting molecules, whereas the number in the denominator only linearly, simplifying to similar terms due to the isotropy of the Green functions. Thus, the superradiance fidelity will increase linearly with the number of molecules and saturate due to the finite molecule--molecule separation and the inverse power law of the interactions. When all particles are excited, the superradiance emission receives a further $N$ leading to the well-known quadratic scaling law. Furthermore, the model can be adapted to studying superradiance in gas clouds, where each molecule can be treated as an effective medium due to its environment~\cite{Aspnes}, leading to local-field corrections~\cite{doi:10.1021/acs.jpca.7b10159}, which can be extended to many molecules and will result in a superradiance fidelity depending on the molecule's density. 

\section*{Acknowledgments}
The authors thank Matthias Bohlen and Frank Stienkemeier for fruitful discussions.
We gratefully acknowledge support from the German Research Council (grants BU 1803/6-1, S.Y.B. and J.F., BU 1803/3-1, S.Y.B.). The computations of this work were carried out on UNINETT Sigma2 high-performance computing resources (grant NN9650K).

\appendix
\section{Dyadic Green functions}
\subsection{For planar interfaces}\label{app:plane}
For large cluster radii, its curvature does not play an important role, and the electromagnetic scattering at the cluster can be approximated by the Fresnel reflection at a planar interface. In this case, the scattering Green function reads~\cite{Buhmann12a}
\begin{eqnarray}
\GGp(\rr,\rr',\omega) = \frac{\mi}{8\pi^2}\int\frac{\md^2 k^\parallel}{k_1^{\perp}}\me^{\mi \kk^\parallel\cdot (\rr-\rr')+\mi k_1^\perp (z+z')}
\left[ r_s \ee_{s+}^1\ee_{s-}^1 +r_p \ee_{p+}^1\ee_{p-}^1\right] \,, \label{eq:GreenPlanar1}
\end{eqnarray}
with the Fresnel reflection coefficients
\begin{eqnarray}
r_s &=& \frac{k_1^\perp -k_2^\perp}{k_1^\perp+k_2^\perp}\,,\\
r_p &=& \frac{\varepsilon_2 k_1^\perp -\varepsilon_1 k_2^\perp}{\varepsilon_2 k_1^\perp+\varepsilon_1 k_2^\perp}\,,
\end{eqnarray}
the wave vector parallel to the plane $\kk^\parallel \perp \ee_z$ and its component towards $z$ direction
\begin{eqnarray}
k^\perp_j = \sqrt{\varepsilon_j \frac{\omega^2}{c^2}-{k^\parallel}^2} \,.
\end{eqnarray}
By introducing spherical coordinates for the $\kk^\parallel$ integral, $\kk=k^\parallel(\cos\varphi,\sin\varphi,0)$, one finds $\ee_{s\pm}^j=\ee_{k^\parallel}\times \ee_z=(\sin\varphi,-\cos\varphi,0)$ and $\ee^j=\frac{1}{k_j}\left(k^\parallel\ee_z\mp k_j^\perp\ee_{k^\parallel}\right)=\frac{c}{\omega\sqrt{\varepsilon_j}}(\mp k_j^\perp\cos\varphi,\mp k_j^\perp\sin\varphi,k^\parallel)$. Thus, the dyads in Eq.~(\ref{eq:GreenPlanar1}) read
\begin{equation}
    \ee_{s+}^1\ee_{s-}^1 = \begin{pmatrix} \sin^2\varphi & -\sin\varphi\cos\varphi&0\\
    -\sin\varphi\cos\varphi& \cos^2\varphi & 0 \\
    0 & 0 & 0
    \end{pmatrix}\,,
\end{equation}
and
\begin{eqnarray}
   \ee_{p+}^1\ee_{p-}^1 =-\frac{c^2}{\omega^2\varepsilon_1}\begin{pmatrix}{k_1^\perp}^2\cos^2\varphi & {k_1^\perp}^2\sin\varphi\cos\varphi & -k^\parallel k_1^\perp \cos\varphi \\
    {k_1^\perp}^2\sin\varphi\cos\varphi & {k^\perp_1}^2\sin^2\varphi & -k^\parallel k_1^\perp \sin\varphi \\
     k^\parallel k_1^\perp \cos\varphi & k^\parallel k_1^\perp \sin\varphi & -{k^\parallel}^2
    \end{pmatrix}\,.
\end{eqnarray}
By choosing the coordinate system such that $\kk^\parallel\cdot(\rr-\rr') =k^\parallel x \cos\varphi$, the $\varphi$ integration can be carried out by using the relations
\begin{equation}
    \int\limits_0^{2\pi}\md \varphi \me^{\mi x\cos\varphi} \cos(n\varphi) = 2\pi \mi^n J_n(x) \,,
\end{equation}
and
\begin{equation}
    \int\limits_0^{2\pi}\md \varphi \me^{\mi x\cos\varphi}\sin(n\varphi) = 0\,,
\end{equation}
with the cylindrical Bessel functions of the first kind $J_n(x)$, the scattering Green function for the planarly layered case can be written as
\begin{eqnarray}
\lefteqn{\GGp(\rr,\rr',\omega)=\frac{\mi}{8\pi}\int\frac{k^\parallel \md k^\parallel}{k_1^\perp} \me^{\mi k_1^\perp(z+z')}}\nonumber\\
&&\times\left[ r_s\begin{pmatrix} J_+(k^\parallel x) & 0 & 0 \\
0 & J_-(k^\parallel x) & 0 \\
0 & 0 & 0
\end{pmatrix}-\frac{r_p c^2}{\omega^2\varepsilon_1}\right.\nonumber\\
&&\left.\times\begin{pmatrix}{k_1^\perp}^2J_-(k^\parallel x) & 0 & 2\mi k^\parallel k_1^\perp J_1(k^\parallel x)\\
0 & {k_1^\perp}^2 J_+(k^\parallel x ) & 0 \\
-2\mi k^\parallel k_1^\perp J_1(k^\parallel x) & 0 & -2 {k^\parallel}^2 J_0(k^\parallel x)
\end{pmatrix}\right]\,,\nonumber\\
\end{eqnarray}
with $J_\pm(k^\parallel x) =J_0(k^\parallel x)\pm J_2(k^\parallel x)$. The coincidence limit is given by $x\mapsto 0$ and $z'=z$, which leads to
\begin{eqnarray}
\GGp(\rr,\rr,\omega)=\frac{\mi}{8\pi}\int\frac{k^\parallel \md k^\parallel}{k_1^\perp} \me^{\mi 2 k_1^\perp z}\left[ r_s\begin{pmatrix} 1 & 0 & 0 \\
0 & 1 & 0 \\
0 & 0 & 0
\end{pmatrix}-\frac{r_p c^2}{\omega^2\varepsilon_1}\begin{pmatrix}{k_1^\perp}^2 & 0 & 0\\
0 & {k_1^\perp}^2  & 0 \\
0 & 0 & -2 {k^\parallel}^2 
\end{pmatrix}\right]\,.
\end{eqnarray}
In the non-retarded limit, which is valid for distances smaller than the relevant wavelength, $z\ll c/\omega_{\rm rel}$, which also means that large values of $k^\parallel$ contribute most to the integral and thus $k_1^\perp =k_2^\perp =\mi k^\parallel$ that further leads to the vanishing of the reflection of $s$-waves, $r_s=0$, the Green function further simplifies to
\begin{eqnarray}
\GGp(\rr,\rr,\omega)=\frac{c^2}{32\pi\omega^2z^3}r_p  \begin{pmatrix}1 & 0 & 0\\
0 & 1  & 0 \\
0 & 0 & 2 
\end{pmatrix}\,.\label{eq:GPLA}
\end{eqnarray}
\subsection{Spherical dyadic Green's function}\label{app:sphere}
The dyadic Green's function in spherical coordinates $(r,\varphi,\vartheta)$ is given by
\begin{eqnarray}
 \lefteqn{{\bf{G}}(\rr,\rr',\omega) = \frac{\mi k_1}{4\pi} \sum_{e,o} \sum_{l=1}^\infty\frac{2l+1}{l(l+1)} \sum_{m=0}^l (2-\delta_{m0})\frac{(l-m)!}{(l+m)!}}\nonumber\\&&\times  \left[ r_s {\bm{M}}_{{}_o^eml}(k_1) {\bm{M}}_{{}_o^eml}'(k_1) + r_p {\bm{N}}_{{}_o^eml}(k_1) {\bm{N}}_{{}_o^eml}'(k_1)\right] \, ,\nonumber\\\label{eq:Green1}
\end{eqnarray}
with spherical vector wave functions
\begin{eqnarray}
{\bm{M}}_{{}_o^eml}(k) &=& \mp\frac{m}{\sin\vartheta}h_l^{(1)}(kr)P_l^m(\cos\vartheta)\begin{matrix}
\sin\\\cos\end{matrix}m\varphi {\bm{e}}_\vartheta-h_l^{(1)}(kr)\frac{\mathrm d P_l^m(\cos\vartheta)}{\mathrm d\vartheta}\begin{matrix}\cos\\\sin\end{matrix} m\varphi {\bm{e}}_\varphi \,,\label{eq:M1}\\ 
{\bm{N}}_{{}_o^eml}(k) &=& \frac{l(l+1)}{kr} h_l^{(1)}(kr)P_l^m(\cos\vartheta)\begin{matrix} \cos\\\sin\end{matrix} m\varphi {\bm{e}}_r\nonumber\\
&&+\frac{1}{kr}\frac{\mathrm d r h_l^{(1)}(kr)}{\mathrm d r} \left[\frac{\mathrm d P_l^m(\cos\vartheta)}{\mathrm d \vartheta}\begin{matrix}\cos\\\sin\end{matrix}m\varphi {\bm{e}}_\vartheta \mp \frac{mP_l^m(\cos\vartheta)}{\sin\vartheta}\begin{matrix}\sin\\\cos\end{matrix}m\varphi {\bm{e}}_\varphi\right]\,, \label{eq:N1}
\end{eqnarray}
with the vacuum wave number $k_1=\omega/c$ and the wave number inside the sphere $k_2=k_1\sqrt{\varepsilon(\omega)}$. The vectors ${\bm{e}}_r$, ${\bm{e}}_\vartheta$ and ${\bm{e}}_\varphi$ are the mutually orthogonal unit vectors. The reflection coefficients for $s$-polarised waves reads as
\begin{equation}
    r_s = -\frac{k_2 \eta_l(k_2 R)j_l(k_1R)-k_1\eta_l(k_1R)j_l(k_2R)}{k_2 \eta_l(k_2 R) h_l^{(1)}(k_1R)-k_1 \zeta_l(k_1 R)j_l(k_2R)} \,, \label{eq:rs1}
\end{equation}
and for $p$-polarised waves as
\begin{equation}
    r_p = -\frac{k_2 \eta_l(k_1R)j_l(k_2R)-k_1\eta_l(k_2R)j_l(k_1R)}{k_2 \zeta_l(k_1R)j_l(k_2R)-k_1\eta_l(k_2R)h_l^{(1)}(k_1R)} \,, \label{eq:rp1}
\end{equation}
with the spherical Bessel and Hankel function of the first kind $j_l(x)$ and $h_l^{(1)}(x)$, respectively, and the Ricatti functions
\begin{equation}
    \eta_l(x) = \frac{1}{x}\frac{\mathrm d xj_l(x)}{\mathrm d x}\,,\qquad \zeta_l(x) = \frac{1}{x}\frac{\mathrm d xh_l^{(1)}(x)}{\mathrm d x}\,.
\end{equation}
In Eq.~(\ref{eq:Green1}) together with the reflection coefficients~(\ref{eq:rs1}) and (\ref{eq:rp1}) it can be observed that the $m$ is only given inside the vector wave functions
\begin{eqnarray}
 {\bf{G}}(\rr,\rr',\omega) &=& \frac{\mi k_1}{4\pi}  \sum_{l=1}^\infty\frac{2l+1}{l(l+1)}   \left[ r_s \sum_{e,o}\sum_{m=0}^l (2-\delta_{m0})  \frac{(l-m)!}{(l+m)!}{\bm{M}}_{{}_o^eml}(k_1) {\bm{M}}_{{}_o^eml}'(k_1)\right.\nonumber\\
 &&\left.+r_p \sum_{e,o}\sum_{m=0}^l (2-\delta_{m0})\frac{(l-m)!}{(l+m)!}{\bm{N}}_{{}_o^eml}(k_1) {\bm{N}}_{{}_o^eml}'(k_1)\right] \,.\label{eq:Green2}
\end{eqnarray}
This leads to the terms

\begingroup
\allowdisplaybreaks
\begin{align}
\lefteqn{{\bf{M}}_l(\rr,\rr',\omega) =\sum_{e,o}\sum_{m=0}^l (2-\delta_{m0})\frac{(l-m)!}{(l+m)!}{\bm{M}}_{{}_o^eml}(k_1) {\bm{M}}_{{}_o^eml}'(k_1)}\nonumber\\=&
h_l^{(1)}(kr)h_l^{(1)}(kr')\sum_{m=0}^l (2-\delta_{m0})\frac{(l-m)!}{(l+m)!}\left\lbrace\left[\frac{m}{\sin\vartheta}P_l^m(\cos\vartheta)\sin m\varphi {\bm{e}}_\vartheta+\frac{\mathrm d P_l^m(\cos\vartheta)}{\mathrm d\vartheta}\cos m\varphi {\bm{e}}_\varphi\right]\right.\nonumber\\
&\times\left. \left[\frac{m}{\sin\vartheta'}P_l^m(\cos\vartheta')\sin m\varphi' {\bm{e}}_{\vartheta'}+\frac{\mathrm d P_l^m(\cos\vartheta')}{\mathrm d\vartheta'}\cos m\varphi' {\bm{e}}_{\varphi'}\right]\right.\nonumber\\
&+\left.\left[\frac{m}{\sin\vartheta}P_l^m(\cos\vartheta)\cos m\varphi {\bm{e}}_\vartheta-\frac{\mathrm d P_l^m(\cos\vartheta)}{\mathrm d\vartheta}\sin m\varphi {\bm{e}}_\varphi\right]\right.\nonumber\\&\left. \left[\frac{m}{\sin\vartheta'}P_l^m(\cos\vartheta')\cos m\varphi' {\bm{e}}_{\vartheta'}-\frac{\mathrm d P_l^m(\cos\vartheta')}{\mathrm d\vartheta'}\sin m\varphi' {\bm{e}}_{\varphi'}\right]\right\rbrace\nonumber\\
=&h_l^{(1)}(kr)h_l^{(1)}(kr')\sum_{m=0}^l (2-\delta_{m0})\frac{(l-m)!}{(l+m)!}\nonumber\\
&\times\left\lbrace \frac{m^2}{\sin\vartheta\sin\vartheta'}P_l^m(\cos\vartheta)P_l^m(\cos\vartheta')\left[\sin m\varphi\sin m\varphi' +\cos m\varphi \cos m \varphi'\right]{\bm{e}}_\vartheta{\bm{e}}_{\vartheta'} \right.\nonumber\\
&\left. +\frac{m}{\sin\vartheta}P_l^m(\cos\vartheta)\frac{\mathrm d P_l^m(\cos\vartheta')}{\mathrm d \vartheta'}\left[\sin m\varphi \cos m\varphi' -\cos m\varphi \sin m\varphi'\right]{\bm{e}}_\vartheta{\bm{e}}_{\varphi'}   \right.\nonumber\\
&\left. +\frac{m}{\sin\vartheta'}P_l^m(\cos\vartheta')\frac{\mathrm d P_l^m(\cos\vartheta)}{\mathrm d \vartheta}\left[\cos m \varphi \sin m\varphi'-\sin m\varphi\cos m\varphi' \right]{\bm{e}}_\varphi{\bm{e}}_{\vartheta'}  \right. \nonumber\\
&+\left.\frac{\mathrm d P_l^m(\cos\vartheta)}{\mathrm d \vartheta}\frac{\mathrm d P_l^m(\cos\vartheta')}{\mathrm d \vartheta'}\left[\cos m\varphi\cos m\varphi'+\sin m\varphi \sin m \varphi'\right]{\bm{e}}_\varphi {\bm{e}}_{\varphi'} \right\rbrace\nonumber\\
=&h_l^{(1)}(kr)h_l^{(1)}(kr')\sum_{m=0}^l (2-\delta_{m0})\frac{(l-m)!}{(l+m)!}\nonumber\\
&\times\left\lbrace \frac{m^2}{\sin\vartheta\sin\vartheta'}P_l^m(\cos\vartheta)P_l^m(\cos\vartheta')\cos\left[m\left(\varphi-\varphi'\right)\right]{\bm{e}}_\vartheta{\bm{e}}_{\vartheta'}\right.\nonumber\\
&\left.+\frac{m}{\sin\vartheta}P_l^m(\cos\vartheta)\frac{\mathrm d P_l^m(\cos\vartheta')}{\mathrm d \vartheta'}\sin\left[ m(\varphi  -\varphi')\right]{\bm{e}}_\vartheta{\bm{e}}_{\varphi'}   \right.\nonumber\\
&\left. -\frac{m}{\sin\vartheta'}P_l^m(\cos\vartheta')\frac{\mathrm d P_l^m(\cos\vartheta)}{\mathrm d \vartheta}\sin\left[ m (\varphi -\varphi') \right]{\bm{e}}_\varphi{\bm{e}}_{\vartheta'}\right.\nonumber\\
&\left.  +\frac{\mathrm d P_l^m(\cos\vartheta)}{\mathrm d \vartheta}\frac{\mathrm d P_l^m(\cos\vartheta')}{\mathrm d \vartheta'}\cos\left[ m(\varphi- \varphi')\right]{\bm{e}}_\varphi {\bm{e}}_{\varphi'} \right\rbrace
\end{align}
\endgroup
These sums can be carried out by using the addition theorem for Legendre polynomials
\begin{eqnarray}
    \lefteqn{P_l(\cos\varphi_1\cos\varphi_2 + \sin\varphi_1\sin\varphi_2\cos\Theta)=}\nonumber\\&&\sum_{m=0}^l (2-\delta_{m0})\frac{(l-m)!}{(l+m)!} P_n^m(\cos\varphi_1)P_n^m(\cos\varphi_2)\cos m \Theta\, .\nonumber\\ \label{eq:additiontheorem1}
\end{eqnarray}
By derivating Eq.~(\ref{eq:additiontheorem1}) twice with respect to $\Theta$, one derives
\begin{eqnarray}
    \lefteqn{-P_l''(\cos\varphi_1\cos\varphi_2 + \sin\varphi_1\sin\varphi_2\cos\Theta)\sin^2\varphi_1\sin^2\varphi_2}\nonumber\\
    &&\times\sin^2\Theta+P_l'(\cos\varphi_1\cos\varphi_2 + \sin\varphi_1\sin\varphi_2\cos\Theta)\nonumber\\
    &&\times\sin\varphi_1\sin\varphi_2\cos\Theta\nonumber\\&=&\sum_{m=0}^l (2-\delta_{m0})\frac{(l-m)!}{(l+m)!} m^2 P_n^m(\cos\varphi_1)P_n^m(\cos\varphi_2)\cos m \Theta\, ,\nonumber\\ \label{eq:additiontheorem21}
\end{eqnarray}
and ones with respect to $\Theta$
\begin{eqnarray}
    \lefteqn{P_l'(\cos\varphi_1\cos\varphi_2 + \sin\varphi_1\sin\varphi_2\cos\Theta)\sin\varphi_1\sin\varphi_2\sin\Theta=}\nonumber\\
    &&\sum_{m=0}^l (2-\delta_{m0})\frac{(l-m)!}{(l+m)!}m P_n^m(\cos\varphi_1)P_n^m(\cos\varphi_2)\sin m \Theta\, .\nonumber\\ \label{eq:additiontheorem11}
\end{eqnarray}
Note that the differentials with respect to $\vartheta$ and $\vartheta'$ commute with the sums. Hence, the dyadic product of the ${\bm{M}}$ vector wave functions simplifies to
\begingroup
\allowdisplaybreaks
\begin{align}
{\bf{M}}_l(\rr,\rr',\omega) &=h_l^{(1)}(kr)h_l^{(1)}(kr')\Biggl\lbrace \left[-P_l''[\cos\vartheta\cos\vartheta' + \sin\vartheta\sin\vartheta'\cos(\varphi-\varphi')]\sin\vartheta\sin\vartheta'\sin^2(\varphi-\varphi')\right.\Biggr.\nonumber\\
&\Biggl.\left.+P_l'[\cos\vartheta\cos\vartheta' + \sin\vartheta\sin\vartheta'\cos(\varphi-\varphi')]\cos(\varphi-\varphi') \right] {\bm{e}}_\vartheta{\bm{e}}_{\vartheta'}  \Biggr.\nonumber\\
&+\frac{\mathrm d P_l'[\cos\vartheta\cos\vartheta' + \sin\vartheta\sin\vartheta'\cos(\varphi-\varphi')]\sin\vartheta'\sin(\varphi-\varphi')}{\mathrm d \vartheta'}{\bm{e}}_\vartheta{\bm{e}}_{\varphi'} \nonumber \\
&-\frac{\mathrm d P_l'[\cos\vartheta\cos\vartheta' + \sin\vartheta\sin\vartheta'\cos(\varphi-\varphi')]\sin\vartheta\sin(\varphi-\varphi')}{\mathrm d \vartheta}{\bm{e}}_\varphi{\bm{e}}_{\vartheta'} \nonumber \\
&\Biggl.+\frac{\mathrm d^2 P_l[\cos\vartheta\cos\vartheta' + \sin\vartheta\sin\vartheta'\cos(\varphi-\varphi')]}{\mathrm d \vartheta\mathrm d \vartheta'}{\bm{e}}_\varphi{\bm{e}}_{\varphi'}\Biggr\rbrace\,.
\end{align}
\endgroup
This equation further simplifies by taking the coincidence limit $\rr'\mapsto\rr$
\begin{eqnarray}
{\bf{M}}_l(\rr,\rr,\omega) &=&\left[h_l^{(1)}(kr)\right]^2 \frac{l(l+1)}{2}\left({\bm{e}}_\vartheta{\bm{e}}_\vartheta+{\bm{e}}_\varphi{\bm{e}}_{\varphi}\right)\,.
\end{eqnarray}
Furthermore, in the limit of both molecules are located on the sphere, $r=R+z$, $\varphi = 0$ and $\vartheta=0$; and $r'=R+z=r$, $\varphi' = 0=\varphi$ and $\vartheta'=\delta/(R+z)$, the $s$-wave scattering simplifies to
\begingroup
\allowdisplaybreaks
\begin{align}
{\bf{M}}_l(\rr,\rr',\omega) &=\left[h_l^{(1)}(kr)\right]^2\Biggl\lbrace P_l'[\cos\vartheta\cos\vartheta' + \sin\vartheta\sin\vartheta']  {\bm{e}}_\vartheta{\bm{e}}_{\vartheta'}  +\frac{\mathrm d^2 P_l[\cos\vartheta\cos\vartheta' + \sin\vartheta\sin\vartheta']}{\mathrm d \vartheta\mathrm d \vartheta'}{\bm{e}}_\varphi{\bm{e}}_{\varphi'}\Biggr\rbrace\\
=&\left[h_l^{(1)}(kr)\right]^2\Biggl\lbrace \frac{(l+1)\left[\cos\vartheta'P_l(\cos\vartheta')-P_{l+1}(\cos\vartheta')\right]}{\sin^2\vartheta'} {\bm{e}}_\vartheta{\bm{e}}_{\vartheta'} \Biggr.\nonumber\\
&\Biggl.+\frac{(l+1)\left[\cos\vartheta' P_{l+1}(\cos\vartheta')-P_l(\cos\vartheta')\left((l+1)\cos^2\vartheta'-l\right)\right]}{\sin^2\vartheta'} {\bm{e}}_\varphi{\bm{e}}_{\varphi'}\Biggr\rbrace\,.
\end{align}
\endgroup

It can be observed that the coincidence limit only depends on the distance and the orientation of the molecule with respect to the surface normal. The same analysis can be performed for the $p$-polarised waves leading to the tensor for the $\bm{N}$ vector wave functions
\begingroup
\allowdisplaybreaks
\begin{align}
\lefteqn{{\bf{N}}_l(\rr,\rr',\omega) = \sum_{e,o}\sum_{m=0}^l (2-\delta_{m0})\frac{(l-m)!}{(l+m)!}{\bm{N}}_{{}_o^eml}(k_1) {\bm{N}}_{{}_o^eml}'(k_1)}\nonumber\\
=& \sum_{e,o}\sum_{m=0}^l (2-\delta_{m0})\frac{(l-m)!}{(l+m)!} \left\lbrace \frac{l(l+1)}{kr} h_l^{(1)}(kr)P_l^m(\cos\vartheta)\begin{matrix} \cos\\\sin\end{matrix} m\varphi {\bm{e}}_r \right.\nonumber\\
&\left.+\frac{1}{kr}\frac{\mathrm d r h_l^{(1)}(kr)}{\mathrm d r} \left[\frac{\mathrm d P_l^m(\cos\vartheta)}{\mathrm d \vartheta}\begin{matrix}\cos\\\sin\end{matrix}m\varphi {\bm{e}}_\vartheta \mp \frac{mP_l^m(\cos\vartheta)}{\sin\vartheta}\begin{matrix}\sin\\\cos\end{matrix}m\varphi {\bm{e}}_\varphi\right]\right\rbrace\left\lbrace \frac{l(l+1)}{kr'} h_l^{(1)}(kr') \right.\nonumber\\
&\left.\times P_l^m(\cos\vartheta')\begin{matrix} \cos\\\sin\end{matrix} m\varphi' {\bm{e}}_{r'}+\frac{h_l^{(1)\prime}(kr')}{kr'} \left[\frac{\mathrm d P_l^m(\cos\vartheta')}{\mathrm d \vartheta'}\begin{matrix}\cos\\\sin\end{matrix}m\varphi' {\bm{e}}_{\vartheta'} \mp \frac{mP_l^m(\cos\vartheta')}{\sin\vartheta'}\begin{matrix}\sin\\\cos\end{matrix}m\varphi' {\bm{e}}_{\varphi'}\right]\right\rbrace \nonumber\\
=&\sum_{m=0}^l (2-\delta_{m0})\frac{(l-m)!}{(l+m)!} \left\lbrace \frac{l^2(l+1)^2}{k^2rr'}h_l^{(1)}(kr)h_l^{(1)}(kr') P_l^m(\cos\vartheta)P_l^m(\cos\vartheta')\right.\nonumber\\
&\left.\times\left[\cos m\varphi \cos m\varphi' +\sin m\varphi \sin m\varphi'\right]{\bm{e}}_r{\bm{e}}_{r'} \right.\nonumber\\
&\left. + \frac{l(l+1)}{k^2rr'} h_l^{(1)}(kr)h_l^{(1)\prime}(kr') P_l^m(\cos\vartheta)\frac{\mathrm d P_l^m(\cos\vartheta')}{\mathrm d \vartheta'}\left[\cos m\varphi \cos m\varphi' +\sin m\varphi \sin m\varphi'\right]{\bm{e}}_r{\bm{e}}_{\vartheta'} \right. \nonumber\\
&\left. + \frac{l(l+1)m}{k^2rr'\sin\vartheta'} h_l^{(1)}(kr)h_l^{(1)\prime}(kr') P_l^m(\cos\vartheta)P_l^m(\cos\vartheta')\left[-\cos m\varphi \sin m\varphi' +\sin m\varphi \cos m\varphi'\right]{\bm{e}}_r{\bm{e}}_{\varphi'}\right.\nonumber\\
&+\left. \frac{l(l+1)}{k^2 rr'}h_l^{(1)\prime}(kr)h_l^{(1)}(kr') \frac{\mathrm d P_l^m(\cos\vartheta)}{\mathrm d \vartheta}P_l^m(\cos\vartheta')\left[\cos m\varphi \cos m \varphi' +\sin m \varphi \sin m\varphi'\right] {\bm{e}}_\vartheta{\bm{e}}_{r'} \right. \nonumber \\
&\left. + \frac{h_l^{(1)\prime}(kr)h_l^{(1)\prime}(kr')}{k^2rr'}\frac{\mathrm d P_l^m(\cos\vartheta)}{\mathrm d \vartheta}\frac{\mathrm d P_l^m(\cos\vartheta')}{\mathrm d \vartheta'}\left[\cos m\varphi\cos m\varphi' +\sin m \varphi \sin m \varphi'\right]{\bm{e}}_\vartheta {\bm{e}}_{\vartheta'}\right.\nonumber \\
&\left.+\frac{h_l^{(1)\prime}(kr)h_l^{(1)\prime}(kr')}{k^2rr'} \frac{m}{\sin\vartheta'}\frac{\mathrm d P_l^m(\cos\vartheta)}{\mathrm d \vartheta}P_l^m(\cos\vartheta')\left[-\cos m \varphi \sin m \varphi' +\sin m\varphi \cos  m\varphi'\right]{\bm{e}}_\vartheta{\bm{e}}_{\varphi'}\right.\nonumber \\
&\left.  \frac{l(l+1)}{k^2 rr'}h_l^{(1)\prime}(kr)h_l^{(1)}(kr')\frac{m}{\sin\vartheta} P_l^m(\cos\vartheta)P_l^m(\cos\vartheta')\left[-\sin m\varphi\cos m\varphi +\cos m \varphi \sin m \varphi'\right]{\bm{e}}_\varphi{\bm{e}}_{r'} \right.\nonumber \\
&\left. +  \frac{h_l^{(1)\prime}(kr)h_l^{(1)\prime}(kr')}{k^2rr'} \frac{m}{\sin\vartheta} P_l^m(\cos\vartheta)\frac{\mathrm d P_l^m(\cos\vartheta')}{\mathrm d \vartheta'} \left[-\sin m\varphi \cos m \varphi' + \cos m \varphi \sin m \varphi'\right]{\bm{e}}_\varphi{\bm{e}}_{\vartheta'} \right.\nonumber\\
&\left. +  \frac{h_l^{(1)\prime}(kr)h_l^{(1)\prime}(kr')}{k^2rr'} \frac{m^2}{\sin\vartheta\sin\vartheta'} P_l^m(\cos\vartheta)P_l^m(\cos\vartheta')\left[\sin m\varphi \sin m\varphi' +\cos m\varphi \cos m \varphi'\right]{\bm{e}}_\varphi{\bm{e}}_{\varphi'} \right\rbrace \nonumber\\
=&\sum_{m=0}^l (2-\delta_{m0})\frac{(l-m)!}{(l+m)!} \left\lbrace \frac{l^2(l+1)^2}{k^2rr'}h_l^{(1)}(kr)h_l^{(1)}(kr') P_l^m(\cos\vartheta)P_l^m(\cos\vartheta')\cos m\left(\varphi -\varphi'\right){\bm{e}}_r{\bm{e}}_{r'} \right.\nonumber\\
&\left. + \frac{l(l+1)}{k^2rr'} h_l^{(1)}(kr)h_l^{(1)\prime}(kr') P_l^m(\cos\vartheta)\frac{\mathrm d P_l^m(\cos\vartheta')}{\mathrm d \vartheta'}\cos m\left(\varphi -\varphi'\right){\bm{e}}_r{\bm{e}}_{\vartheta'} \right. \nonumber\\
&\left. + \frac{l(l+1)m}{k^2rr'\sin\vartheta'} h_l^{(1)}(kr)h_l^{(1)\prime}(kr') P_l^m(\cos\vartheta)P_l^m(\cos\vartheta')\sin m\left(\varphi -\varphi'\right){\bm{e}}_r{\bm{e}}_{\varphi'}\right.\nonumber\\
&+\left. \frac{l(l+1)}{k^2 rr'}h_l^{(1)\prime}(kr)h_l^{(1)}(kr') \frac{\mathrm d P_l^m(\cos\vartheta)}{\mathrm d \vartheta}P_l^m(\cos\vartheta')\cos m\left(\varphi -\varphi'\right) {\bm{e}}_\vartheta{\bm{e}}_{r'} \right. \nonumber \\
&\left. + \frac{h_l^{(1)\prime}(kr)h_l^{(1)\prime}(kr')}{k^2rr'}\frac{\mathrm d P_l^m(\cos\vartheta)}{\mathrm d \vartheta}\frac{\mathrm d P_l^m(\cos\vartheta')}{\mathrm d \vartheta'}\cos m\left(\varphi -\varphi'\right){\bm{e}}_\vartheta {\bm{e}}_{\vartheta'}\right.\nonumber \\
&\left.+\frac{h_l^{(1)\prime}(kr)h_l^{(1)\prime}(kr')}{k^2rr'} \frac{m}{\sin\vartheta'}\frac{\mathrm d P_l^m(\cos\vartheta)}{\mathrm d \vartheta}P_l^m(\cos\vartheta')\sin m\left(\varphi -\varphi'\right){\bm{e}}_\vartheta{\bm{e}}_{\varphi'}\right.\nonumber \\
&\left. - \frac{l(l+1)}{k^2 rr'}h_l^{(1)\prime}(kr)h_l^{(1)}(kr')\frac{m}{\sin\vartheta} P_l^m(\cos\vartheta)P_l^m(\cos\vartheta')\sin m\left(\varphi -\varphi'\right){\bm{e}}_\varphi{\bm{e}}_{r'} \right.\nonumber \\
&\left. -  \frac{h_l^{(1)\prime}(kr)h_l^{(1)\prime}(kr')}{k^2rr'} \frac{m}{\sin\vartheta} P_l^m(\cos\vartheta)\frac{\mathrm d P_l^m(\cos\vartheta')}{\mathrm d \vartheta'} \sin m\left(\varphi -\varphi'\right){\bm{e}}_\varphi{\bm{e}}_{\vartheta'} \right.\nonumber\\
&\left. +  \frac{h_l^{(1)\prime}(kr)h_l^{(1)\prime}(kr')}{k^2rr'} \frac{m^2}{\sin\vartheta\sin\vartheta'} P_l^m(\cos\vartheta)P_l^m(\cos\vartheta')\cos m\left(\varphi -\varphi'\right){\bm{e}}_\varphi{\bm{e}}_{\varphi'} \right\rbrace\,.
\end{align}
\endgroup
Now, the the sum over $m$ can be carried out together with the addition theorem~(\ref{eq:additiontheorem1})--(\ref{eq:additiontheorem11}), which leads to
\begingroup
\allowdisplaybreaks
\begin{align}
    \lefteqn{{\bf{N}}_l(\rr,\rr',\omega) = \Biggl\lbrace \frac{l^2(l+1)^2}{k^2rr'}h_l^{(1)}(kr)h_l^{(1)}(kr') P_l\left[\cos\vartheta\cos\vartheta' + \sin\vartheta\sin\vartheta'\cos \left(\varphi-\varphi'\right)\right]{\bm{e}}_r{\bm{e}}_{r'} \Biggr.}\nonumber\\
&\left. + \frac{l(l+1)}{k^2rr'} h_l^{(1)}(kr)h_l^{(1)\prime}(kr') \frac{\mathrm d P_l\left[\cos\vartheta\cos\vartheta' + \sin\vartheta\sin\vartheta'\cos (\varphi-\varphi')\right]}{\mathrm d \vartheta'}{\bm{e}}_r{\bm{e}}_{\vartheta'} \right. \nonumber\\
&\left. + \frac{l(l+1)}{k^2rr'} h_l^{(1)}(kr)h_l^{(1)\prime}(kr') P_l'\left[\cos\vartheta\cos\vartheta' + \sin\vartheta\sin\vartheta'\cos(\varphi-\varphi')\right]\sin\vartheta\sin\left(\varphi-\varphi'\right){\bm{e}}_r{\bm{e}}_{\varphi'}\right.\nonumber\\
&+\left. \frac{l(l+1)}{k^2 rr'}h_l^{(1)\prime}(kr)h_l^{(1)}(kr') \frac{\mathrm d P_l\left[\cos\vartheta\cos\vartheta' + \sin\vartheta\sin\vartheta'\cos (\varphi-\varphi')\right]}{\mathrm d \vartheta} {\bm{e}}_\vartheta{\bm{e}}_{r'} \right. \nonumber \\
&\left. + \frac{h_l^{(1)\prime}(kr)h_l^{(1)\prime}(kr')}{k^2rr'}\frac{\mathrm d^2 P_l\left[\cos\vartheta\cos\vartheta' + \sin\vartheta\sin\vartheta'\cos (\varphi-\varphi')\right]}{\mathrm d \vartheta\mathrm d \vartheta'}{\bm{e}}_\vartheta {\bm{e}}_{\vartheta'}\right.\nonumber \\
&\left.+\frac{h_l^{(1)\prime}(kr)h_l^{(1)\prime}(kr')}{k^2rr'} \frac{\mathrm d P_l'\left[\cos\vartheta\cos\vartheta' + \sin\vartheta\sin\vartheta'\cos(\varphi-\varphi')\right]\sin\vartheta\sin\left(\varphi-\varphi'\right)}{\mathrm d \vartheta}{\bm{e}}_\vartheta{\bm{e}}_{\varphi'}\right.\nonumber \\
&\left. - \frac{l(l+1)}{k^2 rr'}h_l^{(1)\prime}(kr)h_l^{(1)}(kr')P_l'\left[\cos\vartheta\cos\vartheta' + \sin\vartheta\sin\vartheta'\cos(\varphi-\varphi')\right]\sin\vartheta'\sin\left(\varphi-\varphi'\right){\bm{e}}_\varphi{\bm{e}}_{r'} \right.\nonumber \\
&\left. -  \frac{h_l^{(1)\prime}(kr)h_l^{(1)\prime}(kr')}{k^2rr'}  \frac{\mathrm d P_l'\left[\cos\vartheta\cos\vartheta' + \sin\vartheta\sin\vartheta'\cos(\varphi-\varphi')\right]\sin\vartheta'\sin\left(\varphi-\varphi'\right)}{\mathrm d \vartheta'} {\bm{e}}_\varphi{\bm{e}}_{\vartheta'} \right.\nonumber\\
&\left. +  \frac{h_l^{(1)\prime}(kr)h_l^{(1)\prime}(kr')}{k^2rr'}  \left\lbrace-P_l''\left[\cos\vartheta\cos\vartheta' + \sin\vartheta\sin\vartheta'\cos(\varphi-\varphi')\right]\sin\vartheta\sin\vartheta'\sin^2(\varphi-\varphi')
\right.\right.\nonumber\\
&\Biggl.\left.+P_l'\left[\cos\vartheta\cos\vartheta'   +\sin\vartheta\sin\vartheta'\cos(\varphi-\varphi')\right]\cos(\varphi-\varphi')\right\rbrace{\bm{e}}_\varphi{\bm{e}}_{\varphi'} \Biggr\rbrace\,,
\end{align}
\endgroup
and can be further simplified to in the coincidence limit $\rr'\mapsto\rr$
\begingroup
\allowdisplaybreaks
\begin{align}
{\bf{N}}_l(\rr,\rr,\omega) = \Biggl\lbrace \frac{l^2(l+1)^2}{k^2r^2}\left[h_l^{(1)}(kr)\right]^2 {\bm{e}}_r{\bm{e}}_{r}  + \left[\frac{1}{kr}\frac{\mathrm d r h_l^{(1)}(kr)}{\mathrm d r}\right]^2\frac{l(l+1)}{2}\left({\bm{e}}_\vartheta {\bm{e}}_{\vartheta}+{\bm{e}}_\varphi{\bm{e}}_\varphi\right) \Biggr\rbrace\,.
\end{align}
\endgroup
Furthermore, in the limit of both molecules are located on the sphere, $r=R+z$, $\varphi = 0$ and $\vartheta=0$; and $r'=R+z=r$, $\varphi' = 0=\varphi$ and $\vartheta'=\delta/(R+z)$, the $p$-wave scattering simplifies to
\begingroup
\allowdisplaybreaks
\begin{align}
    \lefteqn{{\bf{N}}_l(\rr,\rr',\omega) = \Biggl\lbrace \frac{l^2(l+1)^2}{k^2r^2}\left[h_l^{(1)}(kr)\right]^2 P_l\cos\vartheta' {\bm{e}}_r{\bm{e}}_{r'} + \frac{l(l+1)}{k^2r^2}\left[h_l^{(1)}(kr)\right]^2 \frac{\mathrm d P_l\left[\cos\vartheta\cos\vartheta' + \sin\vartheta\sin\vartheta'\right]}{\mathrm d \vartheta'}{\bm{e}}_r{\bm{e}}_{\vartheta'}\Biggr.}\nonumber\\
&+\left. \frac{l(l+1)}{k^2 r^2}\left[h_l^{(1)}(kr)\right]^2 \frac{\mathrm d P_l\left[\cos\vartheta\cos\vartheta' + \sin\vartheta\sin\vartheta'\right]}{\mathrm d \vartheta} {\bm{e}}_\vartheta{\bm{e}}_{r'} + \frac{\left[h_l^{(1)\prime}(kr)\right]^2}{k^2r^2}\frac{\mathrm d^2 P_l\left[\cos\vartheta\cos\vartheta' + \sin\vartheta\sin\vartheta'\right]}{\mathrm d \vartheta\mathrm d \vartheta'}{\bm{e}}_\vartheta {\bm{e}}_{\vartheta'}\right. \nonumber \\
&\Biggl.+\frac{\left[h_l^{(1)\prime}(kr)\right]^2}{k^2r^2}P_l'\left[\cos\vartheta\cos\vartheta'   +\sin\vartheta\sin\vartheta'\right]{\bm{e}}_\varphi{\bm{e}}_{\varphi'} \Biggr\rbrace\,,
\end{align}
\endgroup
and can be further simplified to in the coincidence limit $\rr'\mapsto\rr$
\begingroup
\allowdisplaybreaks

To this end, the scattering Green's function in the coincidence limit reads as
\begin{eqnarray}
 {\bf{G}}(\rr,\rr,\omega) &=& \frac{\mi k_1}{8\pi}  \sum_{l=1}^\infty(2l+1)  \Biggl[ r_s \left[h_l^{(1)}(k_1r)\right]^2 \left({\bm{e}}_\vartheta {\bm{e}}_{\vartheta}+{\bm{e}}_\varphi{\bm{e}}_\varphi\right)\Biggr.\nonumber\\&&\Biggl.+ r_p \Biggl\lbrace 2\frac{l(l+1)}{k_1^2r^2}\left[h_l^{(1)}(k_1r)\right]^2 {\bm{e}}_r{\bm{e}}_{r} + \left[\frac{h_l^{(1)\prime}(k_1r)}{k_1r}\right]^2\left({\bm{e}}_\vartheta {\bm{e}}_{\vartheta}+{\bm{e}}_\varphi{\bm{e}}_\varphi\right) \Biggr\rbrace\Biggr] \, .\label{eq:Greenr=r'1}
\end{eqnarray}
The $rr$-component of the spherical Green function reduces to
\begin{eqnarray}
\lefteqn{\GG_{rr}(\rr,\rr',\omega) = \frac{-\mi}{4\pi k_0r^2}\sum_{l=1}^\infty(2l+1)l(l+1) } \nonumber\\
&&\times \frac{\sqrt{\varepsilon}k_0 \eta_l(k_0R)j_l(\sqrt{\varepsilon}k_0R)-k_0\eta_l(\sqrt{\varepsilon}k_0R)j_l(k_0R)}{\sqrt{\varepsilon}k_0 \zeta_l(k_0R)j_l(\sqrt{\varepsilon}k_0R)-k_0\eta_l(\sqrt{\varepsilon}k_0R)h_l^{(1)}(k_0R)}\nonumber\\&&\times \left[h_l^{(1)}(k_0r)\right]^2 P_l\cos\vartheta' \,.
\end{eqnarray}

\section{Rotational average}\label{app:rot}
According to the results of internal dynamics of a quantised molecule system coupled to an environment, the change of transition rates are proportional to
\begin{equation}
    \Gamma_n \propto {\bm{d}}_{nk} \cdot \IM \GG (\rrA,\rrA,\omega_{nk})\cdot {\bm{d}}_{kn}\,.
\end{equation}
The vectorial component of the planar Green function~(\ref{eq:GPLA}) is described by $\ee_x\ee_x+\ee_y\ee_y+2\ee_z\ee_z$. Due to the bounding of the molecule onto the planar surface, the molecule's $z$-component needs to be aligned with the $z$-component of the surface, whereas the remaining components can be orientated arbitrarily to each other. To this end, we average over the remaining orientation
\begin{eqnarray}
    \Gamma_n \propto \frac{1}{2\pi}\int\limits_0^{2\pi}\md\vartheta\, {\bf{R}}(\vartheta) \cdot {\bm{d}}_{nk} \cdot \IM \GG (\rrA,\rrA,\omega_{nk})\cdot{\bf{R}}(\vartheta)\cdot {\bm{d}}_{kn}\,,\nonumber\\
\end{eqnarray}
with the rotation matrix along the $z$-axis
\begin{eqnarray}
 {\bm{R}}(\vartheta) = \begin{pmatrix}
 \cos\vartheta & -\sin\vartheta & 0 \\
 \sin\vartheta & \cos\vartheta & 0 \\
 0 & 0 & 1
 \end{pmatrix}\,,
\end{eqnarray}
which results in
\begin{equation}
    \Gamma_n \propto d_x^2+d_y^2+2d_z^2\,.
\end{equation}
By considering two separated molecules for the superradiance, each molecule has to be averaged for its own
\begin{eqnarray}
     \lefteqn{\Gamma_n \propto \frac{1}{(2\pi)^2}}\nonumber\\&&\times\int\limits_0^{2\pi}\md\vartheta\md\vartheta'\, {\bf{R}}(\vartheta) \cdot {\bm{d}}_{nk} \cdot \IM \GG (\rrA,\rrB,\omega_{nk})\cdot{\bf{R}}(\vartheta')\cdot {\bm{d}}_{kn}\nonumber\\&=&\frac{2}{(2\pi)^2}d_z\,.
\end{eqnarray}
This results can directly be transferred to the spherical case with $\GGs = A_r \ee_r\ee_{r}+A_\vartheta\ee_\vartheta\ee_{\vartheta}+A_\varphi\ee_\varphi\ee_{\varphi}$ by aligning the molecule $z$-axis perpendicular to the surface $\ee_r$, which results in
\begin{equation}
    \Gamma_n \propto A_\varphi \left(d_x^2+ d_y^2\right)+A_rd_z^2\,,
\end{equation}
for the coincidence case. According to the construction of the $s$- and $p$-wave scattering matrices, ${\bf{M}}$ and ${\bf{N}}$, respectively, the vectorial dependence of the Green function can be written as
\begin{eqnarray}
    \lefteqn{\GGs = A_{rr'}\ee_{r}\ee_{r'} + A_{r\vartheta'}\ee_r\ee_{\vartheta'}}\nonumber\\&&+A_{\vartheta r'}\ee_\vartheta\ee_{r'}+A_{\vartheta\vartheta'}\ee_\vartheta\ee_{\vartheta'}+ A_{\varphi\varphi'} \ee_\varphi\ee_{\varphi'}\,,
\end{eqnarray}
which results in
\begin{equation}
    \Gamma_n \propto \frac{A_{rr'}}{(2\pi)^2}d_z^2\,,
\end{equation}
for the rotational average.

\bibliographystyle{unsrt}  
\bibliography{bibi.bib}

\end{document}